\documentclass[conference]{IEEEtran}
\newcommand{\arXiv}{}
\ifdefined\arXiv
\else
    \IEEEoverridecommandlockouts
    \IEEEpubid{\makebox[\columnwidth]{
    ISBN 978-3-903176-28-7 \copyright 2020 IFIP
    \hfill
    } \hspace{\columnsep}\makebox[\columnwidth]{ }}
\fi

\usepackage{url}
\usepackage{amssymb}
\usepackage{float}
\usepackage{balance}
\usepackage[linesnumbered,ruled,noend]{algorithm2e}
\usepackage{amsthm,amstext,amsfonts,bm,amssymb}

\newtheorem{theorem}{Theorem}
\newtheorem{lemma}{Lemma}

\usepackage{pgfplots}
\pgfplotsset{compat=1.14}
\pgfplotsset{
    discard if not/.style 2 args={
        x filter/.code={
            \edef\tempa{\thisrow{#1}}
            \edef\tempb{#2}
            \ifx\tempa\tempb
            \else
                
            \fi
        }
    }
}

\usepackage{subfloat}
\usepackage{mwe}

\ifdefined\arXiv
\else
\newtheorem{innercustomgeneric}{\customgenericname}
\providecommand{\customgenericname}{}
\newcommand{\newcustomtheorem}[2]{%
  \newenvironment{#1}[1]
  {%
   \renewcommand\customgenericname{#2}%
   \renewcommand\theinnercustomgeneric{##1}%
   \innercustomgeneric
  }
  {\endinnercustomgeneric}
}

\newcustomtheorem{customthm}{Theorem}
\newcustomtheorem{customlemma}{Lemma}
\fi
    
\usepackage{graphicx}
\usepackage{framed}
\usepackage{xspace}
\usepackage{amsmath}
\usepackage{graphicx}
\usepackage{mathrsfs}
\usepackage{hhline}
\DeclareMathAlphabet{\mathpzc}{OT1}{pzc}{m}{it}
\usepackage{subfig}

\ifdefined\arXiv
\else
\usepackage[noend]{algpseudocode}

\fi

\newcommand{\eps}{\epsilon}
\newcommand{\set}[1]{\left\{#1\right\}}
\newcommand{\angles}[1]{\left\langle#1\right\rangle}
\newcommand{\brackets}[1]{\left[#1\right]}
\newcommand{\ceil}[1]{ \left\lceil{#1}\right\rceil}

\newcommand{\parentheses}[1]{ \left({#1}\right)}
\newcommand{\abs}[1]{ \left|{#1}\right|}

\providecommand{\tabularnewline}{\\}

\def\_{\,\,\,\,\,}

 \def\Pr{{\mathrm{Pr}}}

\newcommand{\tab}[1]{\hspace{.12\textwidth}\rlap{#1}}

\ifdefined\arXiv
\else
\usepackage[noend]{algpseudocode}
\makeatletter
\newcommand*{\inlineequation}[2][]{%
  \begingroup
    \refstepcounter{equation}%
    \ifx\\#1\\%
    \else
      \label{#1}%
    \fi
    \relpenalty=10000 %
    \binoppenalty=10000 %
    \ensuremath{%
      #2%
    }%
    ~\@eqnnum
  \endgroup
}
\makeatother
\fi
\usepackage{multicol}
\usepackage{multirow}
\usepackage{amssymb}
\usepackage{pifont}

\newcommand{\ignore}[1]{}

\usepackage{xcolor}
\usepackage{array}
\usepackage{enumitem}
\usepackage{comment}
\usepackage{hyperref}

\ifdefined\arXiv
\else
\newcommand{\cutSpaceAfterCaption}{\vspace{-0.043cm}} 
use -0.12cm 
\fi

\newcommand{\sysName}{AROMA\xspace}

\allowdisplaybreaks
\usepackage[]{multirow}

\begin{document}
\title{Routing Oblivious Measurement Analytics}




\author{
 Ran Ben Basat           \\ \small Harvard University
\and Xiaoqi Chen         \\ \small Princeton University
\and Gil Einziger        \\ \small Ben Gurion University 
\and Shir Landau Feibish \\ \small Princeton University
\and Danny Raz           \\ \small Technion
\and Minlan Yu           \\ \small Harvard University
}

%
%
%
%

\allowdisplaybreaks
\renewcommand{\epsilon}{\varepsilon}


\maketitle
\begin{abstract}
Network-wide traffic analytics are often needed for various network monitoring tasks. These measurements are often performed by collecting samples at network switches, which are then sent to the controller for aggregation.
However, performing such analytics without ``overcounting'' flows or packets that traverse multiple measurement switches is challenging. Therefore, existing solutions often simplify the problem by making assumptions on the routing or measurement switch placement.

We introduce AROMA, a measurement infrastructure that generates a uniform sample of packets and flows regardless of the topology, workload and routing. Therefore, AROMA can be deployed in many settings, and can also work in the data plane using programmable PISA switches.
The AROMA infrastructure includes controller algorithms that approximate a variety of essential measurement tasks while providing formal accuracy guarantees.  Using extensive simulations on real-world network traces, we show that our algorithms are competitively accurate compared to the best existing solutions despite the fact that they make no assumptions on the underlying network or the placement of measurement switches.
\end{abstract}


\section{Introduction}\label{sec:intro}

Many network applications, such as load balancing, QoS enforcement, intrusion detection, and traffic engineering~\cite{SONATA,HHHConext,TrafficEngeneering,8888046,LoadBalancing,IntrusionDetection2,Pingmesh,ApproximateFairness,Trumpet,IntrusionDetection}, rely on network-wide analytics of the network traffic to reach informed decisions. 
Network-wide analytics is often done by collecting samples of traffic from individual switches~\cite{facebookDC}. These samples are then sent to the controller, where the data from all measurement switches is combined to assemble a network-wide view of the traffic. Such samples can approximate a variety of essential measurement tasks such as identifying the heavy hitter flows~\cite{RAP,Intervals}, calculating hierarchical heavy hitters~\cite{RHHH,zhang2004online}, estimating the flow size distribution, and identifying super-spreaders \mbox{and port scans~\cite{PortScanners,SuperSpreaders}.}

Samples may be collected at either \emph{flow} or \emph{packet} granularity.
In \emph{packet sampling}, each measurement device samples a packet with probability $p$. 
\emph{Flow sampling}~\cite{flowSampConstraints,cSamp,FlowSampling} is a more complex primitive, where each measurement device samples packets based on their flows; however all flows in the network should be sampled with \emph{the same} probability, regardless of their size.
Inherently, packets that traverse multiple devices, or heavier flows,  are more likely to be sampled, therefore skewing the network-wide measurements. This raises the need for uniform sampling techniques.
In uniform sampling, any packet has an equal chance of being included in the globally collected sample regardless of the number of measurement switches it goes through (as long as it traverses at least one measurement switch). Similarly, any flow has an equal chance of being sampled regardless of the number of packets in the flow. 

To achieve uniform sampling, existing works often assume that packets only traverse a single device~\cite{NetworkWide,zhao2010global}. However, this restricts the measurement device positioning and may only work for specific routing protocols and network topology (e.g., fat trees).
Other solutions rely on packet marking to ensure that each packet is considered only once~\cite{HHTagging}. This technique can be easily exploited by adversaries (e.g., an attacker can mark all its packets to avoid detection). 
Alternatively, some solutions make non-trivial assumptions abut the underlying network characteristics, such as network topology, routing protocols, and traffic dynamics.
For example, cSamp~\cite{cSamp} requires per-switch configuration, and a traffic matrix detailing the number of flows between each source and destination. While these requirements may be sustainable for some networks, they may not be reasonable in large, rapidly changing networks. 

In this paper, we present AROMA  
(\textbf{A}pproximate 
\textbf{R}outing 
\textbf{O}blivious 
\textbf{M}easurement 
\textbf{A}nalytics),
a measurement analytics infrastructure that collects uniform flow and packet samples in any network topology. AROMA is workload and routing oblivious, so packets can traverse multiple measurement devices without biasing the sample.
The underlying technique used in AROMA is a $k$-partition hash based structure, which supports sampling based on the packet or flow identifier.
Therefore, AROMA can be configured to perform packet sampling, flow sampling, or both. 
%
The main \mbox{features of AROMA are:}
\textbf{Routing and workload oblivious measurements. }
AROMA makes no assumptions about the network topology or routing and makes no packet modifications.
Measurement switches produce samples without any need for coordination between them (e.g., there is no need to tag a packet\cite{HHTagging}) and without per-switch configuration (i.e., there is no need to divide responsibility for sampling parts of the traffic across the measurement switches~\cite{cSamp}). 
Such per-switch configurations require prior knowledge about routing and traffic distribution. Acquiring this information incurs substantial overheads in setting up the measurement infrastructure. Furthermore, network and traffic dynamics require continually updating these configurations, which further complicates the solution. Our method avoids such overheads and requires no information on routing, workload, or network topology. It is also the first to perform flow sampling \mbox{without such information.}
\\
\textbf{Supporting a wide range of measurements.}
AROMA supports numerous controller algorithms that utilize the packet and flow samples to accomplish a variety of network measurement tasks, such as estimating the number of (different) packets and flows in the measurement, estimating per-flow frequency, identifying the heavy hitter flows, calculating
\ifdefined\arXiv
\vbox{\noindent hierarchical heavy hitters, estimating the flow size distribution, and identifying super-spreaders.}
\else
hierarchical heavy hitters, estimating the flow size distribution, and identifying super-spreaders. 
\fi

\textbf{PISA compatible.} AROMA can be implemented using P4 and deployed on PISA (Protocol Independent Switch Architecture) programmable switches. Furthermore, the overall switch resources required by AROMA are minimal, which allows the switch to perform additional functionality, giving AROMA a distinct advantage over \mbox{existing monitoring techniques~\cite{PRECISION,FlowRadar,UnivMon,HashPipe}.}

\textbf{Accurate network-wide measurements.}
Evaluation on real-world network traces shows AROMA provides accurate measurements for a variety of network tasks.  Additionally, it is close in performance to existing solutions that provide similar guarantees, but  cannot be implemented using the limited resources and functionality of PISA switches.
For example, with just 0.5MB of space on a trace of 32 million packets,
AROMA estimates flow sizes with a root mean square error of just 150 packets,
achieves an F1 score of 0.9 in identifying superspreaders and of 0.8 in finding heavy hitters,
and estimates the flow size distribution with a weighted mean relative difference of just 0.045.

\section{The \sysName Framework}\label{sec:algs}


  In a nutshell, 
  \sysName is partitioned into a data plane module for sample collection and a controller for  measurement analysis.
  The data plane module collects and maintains packet and flow samples on individual programmable switches, while the controller merges the samples across switches to form a uniform network-wide sample.
  \sysName guarantees that all packets (or flows) have an equal chance to be sampled, as long as it passes through at least one measurement switch.  As a result, our system is routing oblivious and its output is mathematically identical regardless of network topology and routing.
  With the combination of packet sampling and flow sampling (as in~\cite{cSamp,FlowSampling}), \sysName supports a large variety of network monitoring tasks.

Specifically, \sysName uses a two phase hash-based sampling technique to first select a sample slot, and each slot retains the element with minimal hash value.  Here, an element is either a packet (in packet sampling) or a flow (in flow sampling). 
The controller then merges the sampling slots from all switches, and attains for each slot the element whose hash value is (globally) minimal. That is, an element is sampled only if its hash value is globally minimal for its corresponding slot. 
Finally, \sysName runs on P4 programmable switches, and naturally fits within the switch constraints. 
Furthermore, our solution requires a minimal number (2-3)
of pipeline stages and can be configured to run with any amount of memory. This allows \sysName to operate alongside higher-level applications such as load balancing or attack detection. 

We first formally define our model, assumptions, and the notations  (Section~\ref{section:prel}). 
Next, we 
show a method to store and collect a uniform sample (of flows, or packets) within the data plane using PISA programmable switches (Section~\ref{sec:dataplaneAlg}).  Subsequently, we present the controller algorithm to merge the samples collected distributively into a network-wide uniform sample, and survey ways to utilize the samples to perform various measurement tasks (Section~\ref{sec:controlAlg}). 

\begin{table}[h]
\ifdefined\arXiv
\else
\Small
\fi
	\centering{
	\begin{tabular}{|c|p{0.8\columnwidth}|}
		
		\hline
		\textbf{\small  Symbol} & \textbf{\small Definition} \tabularnewline
		\hline
		$\mathcal S$ & The packet stream \tabularnewline
		\hline
		$\angles{\mathit{fid_i},\mathit{pid_i}}$ & A packet from flow $\mathit{fid_i}$ and packet identifier $\mathit{pid_i}$ \tabularnewline
		\hline		
		$\mathcal U$ & The universe of flow identifiers \tabularnewline
		\hline		
		$f_x$ & The frequency of flow $x\in\mathcal U$ \tabularnewline
		\hline				
		$\widehat{V}$ & an estimate for $|\mathcal S|$\tabularnewline
		\hline						
		$\eps$ & The goal error parameter \tabularnewline
		\hline							
		$\delta$ & The goal error probability \tabularnewline
		\hline										
		\multirow{1}{*}{$M$} & The number of samples required for the accuracy guarantee \tabularnewline
		\hline
		\multirow{2}{*}{$\alpha$} & A space factor that allows faster convergence ($\alpha\ge 1$). We use $\alpha\cdot M$ slots instead of $M$. \tabularnewline
		\hline
		\multirow{2}{*}{$\widetilde{M}$}& The number of samples the algorithm produced ($\widetilde{M}\in[0,\alpha\cdot M])$ \tabularnewline
		\hline
		$\widehat{p}$ & The estimated sampling probability ($\widehat{p}=\widetilde{M}/\widehat{V}$) \tabularnewline
		\hline
		$T$ & The actual sample produced ($|T|=\widetilde{M})$ \tabularnewline
		\hline
		$T_x$ & The number of times $x$ appears in the sample $T$ \tabularnewline
		\hline
		\multirow{1}{*}{$\widehat{{f_x}}$} & An estimate for the frequency of flow $x$ (i.e., \quad{}$\widehat{{f_x}}=T_x /  \widehat{p}$)\tabularnewline
		\hline
		$\theta$ &Heavy hitter threshold\tabularnewline
		\hline
		\multirow{1}{*}{$\widehat{M}(t)$} & The number of samples the algorithm produced by time $t$ \tabularnewline
		\hline		
	\end{tabular}
	\ifdefined\arXiv
    \else
	\vspace{0.01cm}
	\caption{List of symbols and notations.}
	\cutSpaceAfterCaption
	\fi
}
	
	\label{table:notations}
\end{table}
\normalsize 

\subsection{Preliminaries}\label{section:prel}
We model the traffic as an ordered \emph{stream} of packets $\mathcal S \in \parentheses{\mathcal{U}\times\mathbb N}^*$, where each packet $\langle \mathit{fid_i}, \mathit{pid_i}\rangle$  has a unique \emph{flow identifier} $\mathit{fid_i}\in \mathcal U$, and a \emph{packet identifier}  $\mathit{pid_i}\in\mathbb N$ that matches each packet to a single flow. 
Flow identifiers can be source IPs, source and destination IP pairs, or $5$-tuples. 
For packet identifiers, in the case of TCP, we can use the TCP sequence number as part of a unique packet identifier. In general, the works of~\cite{TrajectorySampling,Everflow} explain how the packet header fields can be used to derive unique packet identifiers. In this work, we assume the existence of unique packet identifiers. 

We use $K$ \emph{measurement switches} $R_1,\ldots R_K$, 
and assume that each packet traverses  \emph{at least one} measurement switch. Yet, some packets may traverse multiple measurement switches, and the routing rules may change during the measurement.
Formally, our only assumption is that suppose each 
switch sees a subset of the stream ($\mathcal S_k\subseteq \mathcal S$), then all the  
switches together cover all the packets $\cup_{k=1}^K \mathcal S_k = \mathcal S$.  

Our model is more general than that of other network-wide measurement solutions that assume that packets only visit a single measurement switch~\cite{NetworkWide,SketchVisor,DHH1}, or each flow is routed through a single fixed path~\cite{FlowRadar}. 
The same model is also used in related work~\cite{HHTagging,NetworkWideANCS} in a software context.

The term flow refers to the set of packets that share the same flow identifier. Given a flow identifier $x$, its \emph{frequency} $f_x$ 
is the number of packets with $x$ as their flow identifier, i.e., $f_x=\left|\{i|\mathit{fid_i}=x\}\right|$.

\subsection{Data Plane Sampling Module}\label{sec:dataplaneAlg}

We now introduce our data plane sampling infrastructure. Section~\ref{overview} provides a high-level overview of the algorithm, while Section~\ref{implementation} provides the P4 implementation details, adhering to the PISA architecture.

\begin{figure}
    \centering
    \includegraphics[width=0.4\textwidth]{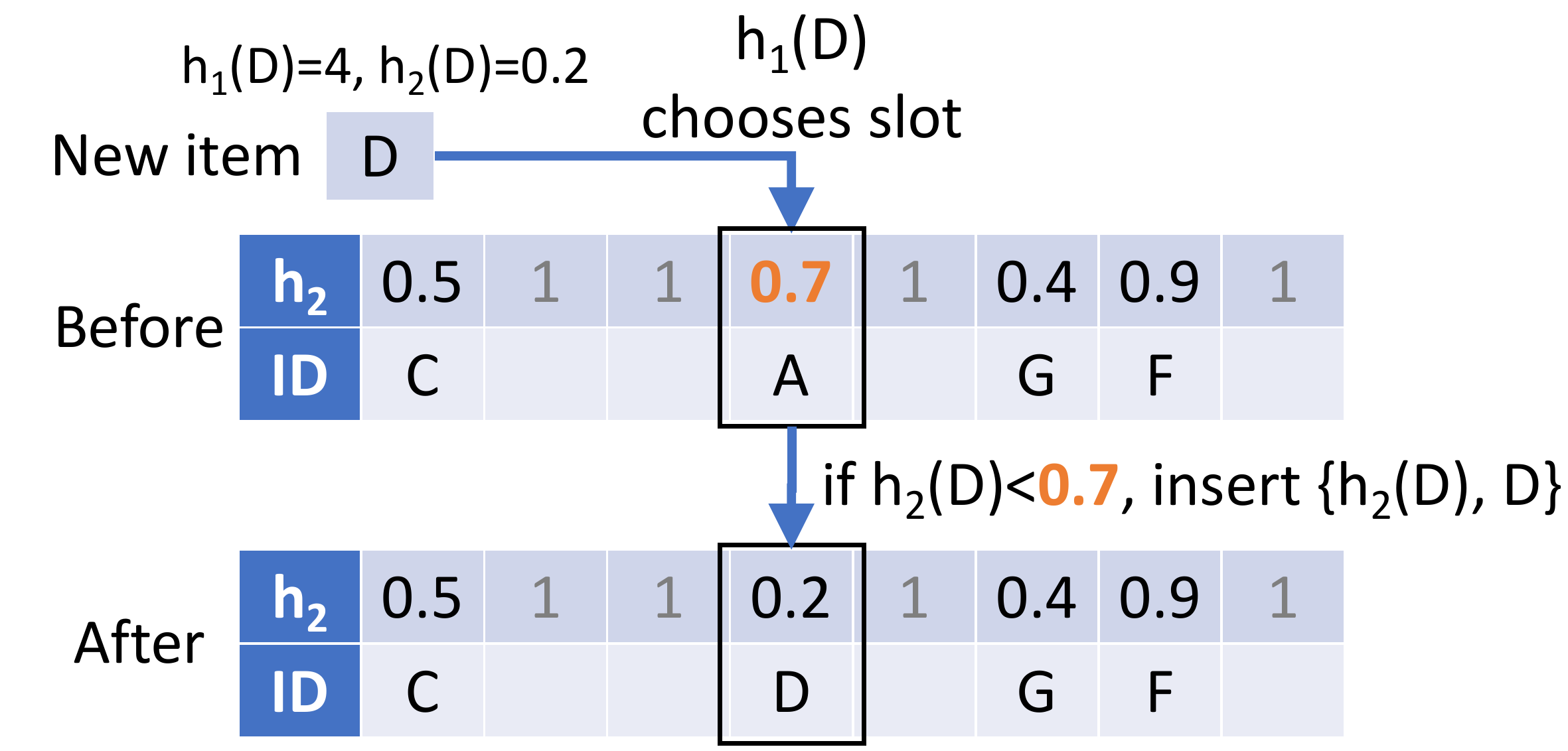}
    \vspace{0.1cm}
    \caption{
    We compute two hashes for each observed item (packet or flow). $h_1$ determines in which slot to compete; if the slot is empty we add the new item. Otherwise, we add it only if its  $h_2$ value is smaller than that of the stored item. 
    }
    \label{fig:DS_insert}
    \ifdefined\arXiv
    \else
    \cutSpaceAfterCaption
    \fi
\end{figure}

\subsubsection{Algorithm overview}\label{overview}
Each measurement switch allocates a fixed size block of memory for $(\alpha \cdot M)$ slots. Measurement switches calculate a hash value in $(0,1]$ based on the packet (or flow) identifier. Each slot stores the item (packet/flow) with the minimal hash value from all the items that were assigned to the slot. Interestingly, hash collisions mean that we require time before the slots are filled. We show that the number of filled slots behaves like a variant of the Coupon Collector problem, but instead of trying to fill \textbf{all} the slots like the common analysis, we attempt to fill a certain percentage of all slots (say 90\%). This relaxation asymptotically reduces the time required to collect the sample at the expense of a slightly inflated memory consumption. 

Formally, each measurement switch observes a stream of packets $\angles{\mathit{fid_i},\mathit{pid_i}}$, with a flow identifier $\mathit{fid_i}$ and packet identifier $\mathit{pid_i}$. We denote $x_i$ as the identifier used, i.e., $x_i=\mathit{fid_i}$ for flow sampling and $x_i=\mathit{pid_i}$ for packet sampling.

Each switch maintains a data structure $MEM$, which contains 
$(\alpha \cdot M)$ memory slots, and each slot stores exactly one identifier. 
The value $\alpha \geq 1$ is selected to ensure a sample of size at least $M$. We provide an analysis of the convergence time in Section~\ref{sec:analysis}.
Let us denote memory slot $j$ as $MEM[j]$. Two values are maintained within the slot: a hash value $MEM[j].hash$ and an identifier $MEM[j].id$.



We use two independent random hash functions: $h_1: \mathcal{U}\rightarrow [0,\alpha \cdot M)$ for mapping an identifier to a memory slot, and
 $h_2: \mathcal{U}\rightarrow (0,1]$ to decide which item to sample. Imperatively, the hash functions are \emph{identical} for the measurement switches that participate in the measurement as it allows merging the data structures for obtaining a network-wide uniform sample and thereby a global view.
 At a high level, each slot receives a fraction of incoming packets, and stores a single identifier $x$ that has the smallest $h_2(x)$ of all those observed by that memory slot. 
 We assume that all measurement switches use \emph{the same} hash functions 
 and that these are chosen at random. This can be done by having the controller randomly choose a seed for pseudo random generator uniformly for all switches at the beginning of the computation epoch.

We initialize all $MEM[j].hash$ to $1$. As illustrated in Figure~\ref{fig:DS_insert}, upon observing each packet and determining identifier $x_i=\texttt{"D"}$, 
 the switch does the following:
 \begin{enumerate}[leftmargin=*]
     \item Computes the two hash values $h_1(x_i)$ and $h_2(x_i)$ for the current packet. In the example above, $h_1(x_i)$ is the fourth {column}, and $h_2(x_i)=0.2$.
     \item Looks up the hash value stored in $MEM[h_1(x_i)].hash$, and ignores the packet if $MEM[h_1(x_i)].hash \leq h_2(x_i)$. 
     \item Otherwise, if $h_2(x_i)<MEM[h_1(x_i)].hash$, then we replace the existing sample in the slot: 
     $$
     \begin{array}{lcl}
          MEM[h_1(x_i)].hash & \leftarrow &h_2(x_i) \\
          MEM[h_1(x_i)].id & \leftarrow &x_i \\
     \end{array}
     $$
 \end{enumerate}

We want to run two instances of our sampling algorithm simultaneously, one for packet-sampling and the other for flow-sampling. Recall that we select $x_i=\mathit{pid_i}$ to sample packets, and $x_i=\mathit{fid_i}$ to sample flows. 


For correctness and accuracy guarantees, we require that at least $M$ out of the $\alpha\cdot M$ slots will not be empty to obtain an $M$-sized uniform sample. We can choose  $\alpha\ge 1$ to expedite this process; in practice, a choice of $\alpha=1.5\sim 2$ suffices. 


\subsubsection{Implementation on PISA programmable switches}\label{implementation}


To achieve Tbps-level aggregated throughput and low forwarding latency, a PISA~\cite{RMT} programmable switch uses a packet processing pipeline architecture that allows only simple operations per pipeline stage, and only has a certain number of hardware stages. 
The work of~\cite{PRECISION} summarized the limitations imposed by PISA switches. 
Most relevant to our case are the limited number of programmable pipeline stages, the  limitations on memory access, and the limitations on arithmetic operations.
\sysName's P4 implementation requires $O(1)$ memory accesses per packet, and can be implemented using only 2 pipeline stages. Thus, it leaves plenty of room for the measurement switch to run other network applications. 



\AlgoDisplayBlockMarkers
\AlgoDisplayGroupMarkers
\SetAlgoBlockMarkers{ }{ \}\ }%
\SetAlgoNoEnd
\SetStartEndCondition{ (}{)}{)}
\SetKwIF{If}{ElseIf}{Else}{if}{\{}{elif}{else\{}{}%
\SetKwFor{While}{while}{\{}{}%
\SetKwBlock{Apply}{apply \{ }{}{}%
\SetKwComment{Comment}{//}{}
\SetKwFunction{Control}{control SampleFlow}

\SetKwBlock{Block}{}{}{}%
  \begin{algorithm}[t]
  \small
  \FuncSty{control FlowSampling(}
      \quad\ArgSty{inout headers} {hdr},
    \tab\tab\tab\tab\quad\quad\quad\quad\quad\ArgSty{inout metadata} {meta}\FuncSty{)} \KwSty{\ \{\ }
 \Block{

     \KwSty{register}$<$ \KwSty{bit}$<$32$>$ $>$ (1$\ll$m)  MEM\textunderscore hash\;  \label{algln:regdefn1}
     
     \KwSty{register}$<$ \KwSty{bit}$<$64$>$ $>$ (1$\ll$m)  MEM\textunderscore id\;  \label{algln:regdefn2}
      \Apply{
      \Comment{Prepare flow identifier $x$}
      
        meta.flowid[31: 0]=hdr.ipv4.srcAddr\;\label{algln:fid1}
      
        meta.flowid[63:32]=hdr.ipv4.dstAddr\;\label{algln:fid2}
      \Comment{Compute hash $h_1(x)\in[0,2^m)$, $h_2(x)\in[0,2^{32})$}
      
        \KwSty{hash}(meta.h1, HashAlgorithm.crc32, 0, \{meta.flowid\}, 1$\ll$m)\;\label{algln:h1}
      
        \KwSty{hash}(meta.h2, HashAlgorithm.crc32\textunderscore custom, 0, \{meta.flowid\}, 1$\ll$32)\;\label{algln:h2}
        
        \vskip 0.2cm
        
        \KwSty{bit}$<$32$>$ existing\textunderscore sample\textunderscore h2\;
          MEM\textunderscore hash.\KwSty{read}(existing\textunderscore sample\textunderscore h2, meta.h1)\;
      \Comment{If the current packet has smaller hash, replace existing sample}
      
        \If{meta.h2$<$existing\textunderscore sample\textunderscore h2}{\label{algln:if}
            MEM\textunderscore hash.\KwSty{write}(meta.h1, meta.h2)\;\label{algln:write1}
            MEM\textunderscore id.\KwSty{write}(meta.h1, meta.flowid)\;\label{algln:write2}
        }      }     }
    \caption{Maintaining uniform flow samples}\label{alg:sampling}
  \end{algorithm}

We now discuss some of the implementation details for performing uniform sampling in the data plane.
At each programmable switch, we allocate two register memory arrays each with $\alpha \cdot M$ entries.  Note that we select $\alpha$ such that $\alpha \cdot M=2^m$ for some $m\in\mathbb{N}$. Such a selection simplifies the implementation as we only have access to random bits, and thus randomizing a number in an arbitrary range is more difficult to implement, and incurs \mbox{additional overheads.}

We denote the register arrays as $MEM^{hash}[i]$ and $MEM^{id}[i]$, and store them in adjacent pipeline stages. The power-of-two sizing of the arrays allows easy addressing using an $m$-bit hash function $h_1$.

Since each array entry on hardware switches usually stores 32 bits,
we store $h_2(x)\in (0,1]$ into $MEM^{hash}$ using 32-bit fixed point encoding.
Hash functions $h_1$ and $h_2$ are implemented using CRC32 with different polynomials, and $h_1$ is truncated to $m$ bits. 
We also initialize all entries in $MEM^{hash}$ to 1. 

As demonstrated by the P4 code shown in Algorithm~\ref{alg:sampling},
for each incoming packet $\angles{\mathit{fid_i},\mathit{pid_i}}$, the programmable switch determines $x_i$ ($x_i \in \{\mathit{fid_i}, \mathit{pid_i}\}$) and does the following:
\begin{enumerate}[leftmargin=*]
    \item 
    Access parsed header fields, such as IPv4 source and destination addresses, to retrieve $x_i$ (line~\ref{algln:fid1},\ref{algln:fid2}), then  compute $h_1(x_i) \in [0,2^m)$ and $h_2(x_i) \in (0,1]$ (line~\ref{algln:h1},\ref{algln:h2}).
    \item
    Compare the value found in $MEM^{hash}[h_1(x_i)]$ to $h_2(x_i)$.
    \begin{itemize}[leftmargin=*]
        \item If $MEM^{hash}[h_1(x_i)]<=h_2(x_i)$: do nothing.
        \item If $MEM^{hash}[h_1(x_i)]>h_2(x_i)$, replace the existing entry with the following as shown in line~\ref{algln:write1} and \ref{algln:write2}:
        \\
        \quad\quad$\begin{array}{lcl}
             MEM^{hash}[h_1(x_i)] & \leftarrow &h_2(x_i), \\
             MEM^{id}[h_1(x_i)] & \leftarrow & x_i, \\
        \end{array}$
        \\
    \end{itemize}
\end{enumerate}

{
 The algorithm uses minimal resources; it computes two hash functions, and resides within two hardware pipeline stages.}
Thus, we can simultaneously implement both flow-sampling and packet-sampling in existing programmable 
switch targets.


    

\subsection{Using the Samples in Control Plane}\label{sec:controlAlg}

The controller merges samples collected from all switches to form a global uniform sample set as described in Section~\ref{sec:merge}. In the subsequent sections, we briefly describe how the controller uses the sample set to perform various measurement tasks. 

\subsubsection{Merging samples}
\label{sec:merge}

\begin{figure}
    \centering
    \includegraphics[width=0.4\textwidth]{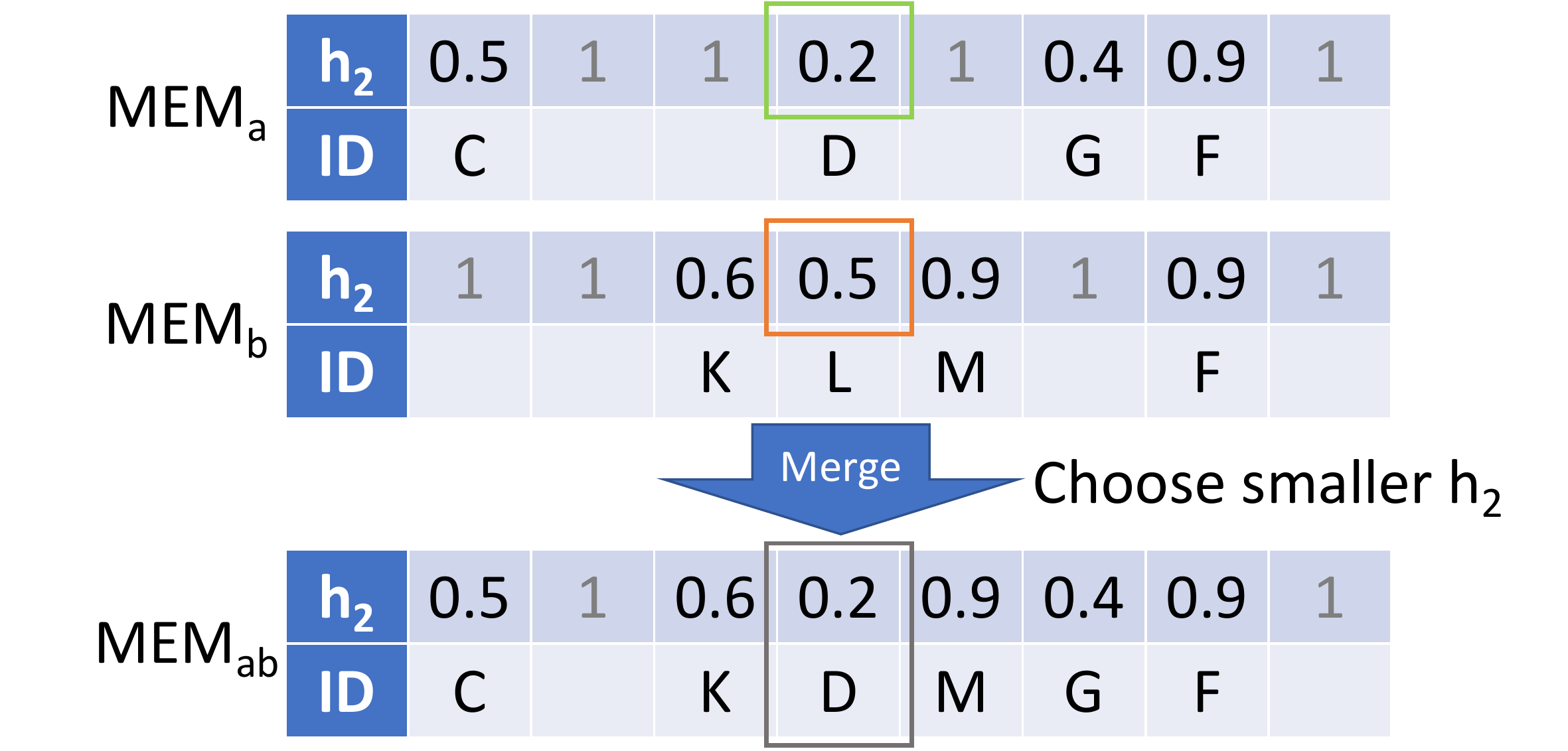}
    \vspace{0.1cm}
    \caption{ \label{fig:DS_merge}
    An example of the merge process of two samples collected in different measurement switches. When the same slot contains different items, we select the one whose $h_2$ value is smaller. In this example, we select ${D}$ \mbox{and discard ${L}$.}
    }
    \ifdefined\arXiv
    \else
   \cutSpaceAfterCaption
   \vspace{-2mm}
   \fi
\end{figure}

First, we describe how to merge two samples into a single sample, as illustrated in Figure~\ref{fig:DS_merge}. Repeatedly applying this algorithm allows the controller to \mbox{merge all the samples.}

Given the samples collected by two switches, $MEM_a[\cdot]$ and $MEM_b[\cdot]$, \sysName merges them to $MEM_{a\cup b}[\cdot]$ as follows: it iterates over the $\alpha \cdot M$ slots, and for each slot $j$ it compares $MEM_a[j].hash$ and $MEM_b[j].hash$ to select the smaller of the two values as the new value for $MEM_{a\cup b}[j].hash$. It then sets $MEM_{a\cup b}[j].id$ accordingly.


It is straightforward to prove that the resulting values in $MEM_{a\cup b}$ are the same as if all packets were observed by either switch $a$ or $b$ or both, as the smaller $h_2$ hash value in each slot will prevail. 
We repeat this process to merge the samples collected at all the switches, to obtain the global sample $MEM_{\mathit{global}}[j].\{hash,id\}$. 
We further trim $MEM_{\mathit{global}}[\cdot]$ to ignore empty slots.





\subsubsection{Number of packets/flows}
Perhaps the most fundamental measurement task is to estimate the actual number of packets and flows within the measurement. 
In the routing oblivious setting, this is not equivalent to summing the number of flows or packets over the different measurement switches as we do not know how many measurement switches each of them traversed.
In the HyperLogLog algorithm~\cite{HLL}, the stream is partitioned into separate substreams, and an independent estimator is maintained for each substream. Each estimator maintains the longest run of consecutive leading zeros of the randomly hashed output of items in its substream. The estimator than uses this information to determine the number of distinct items in each substream.
Similarly to HyperLogLog, by looking at the hash value in each slot, we can infer how many distinct identifiers contested for that slot.


More precisely, given the hash value stored in 
the $i$'th slot
is $MEM_{\mathit{global}}[i].hash \in (0,1]$, the expected total number of distinct identifiers hashed to this slot is $\frac{1}{MEM_{\mathit{global}}[i].hash}$. Thus, by scaling up the harmonic mean of each slot's estimate, the total number of distinct identifiers seen by all the slots can be estimated by:
$
\widehat{V} = \frac{(\alpha\cdot M)^2}{\sum_{i=0}^{\alpha \cdot M-1} (MEM_{\mathit{global}}[i].hash)}.
$




Given this estimated number of different packets/flows, we also get an estimation of the sampling probability. 
For that, assume that $\widetilde{M}\in[0,\alpha\cdot M]$ slots were filled (i.e., we have a uniform sample of size $\widetilde{M}$); then the \emph{estimated} sampling probability is
$\widehat{p} = {\widetilde{M}}/{\widehat{V}}.$
We require the estimated sampling probability for other measurement tasks (such as frequency estimation, superspreaders, and frequency \mbox{distribution estimation).}

\subsubsection{Distributed frequency estimation}

To estimate flow size $f_x$ for a flow $x$, we inspect the uniform packet sample set and look at the packet identifiers. We denote by
$T_x=|\{0\leq i\leq \alpha \cdot M\ |\ MEM_{global}[i].id \in x\}|$  the number of packets in the global uniform sample set that belong to flow $x$. 
Subsequently, we divide $T_x$ by the estimated sampling probability $\widehat{p}$ to get estimated flow size 
$
\widehat{f_x}=T_x / \widehat{p}.
$


\subsubsection{Distributed heavy hitters}
We can use the uniformly sampled packets to estimate heavy hitters, defined as those flows with size $f_x$ which exceeds a $\theta$ fraction of total packet traffic ($|\mathcal S|$), i.e., $f_x >\theta \cdot |\mathcal S|$.
Our algorithm outputs every flow whose frequency in the sample is at least a $\theta$-fraction of the sample size. For example, if $\theta=1\%$ and we gathered $\widetilde{M}=10000$ samples, we will output every flow that appears in the sample at least $100$ times. If an application is more recall-oriented or precision-oriented it is possible to change the threshold to get (with high probability) 100\% accuracy in one of them (at the cost of degrading the other).



\subsubsection{Hierarchical heavy hitters}
We look at the uniformly sampled packets to determine hierarchical heavy hitters. We report a prefix as a hierarchical heavy hitter if it appears in more than $\theta \cdot \widetilde{M}$ packets. 



\subsubsection{Superspreaders} 
We define a Superspreader as a source IP address that communicates with more than $\Psi$ destination IP addresses. 
Such an IP address appear in many flows and is therefore likely to appear in the uniform flow sample.
Given a uniform sample of $\widetilde{M}$ flows,
we can examine the flow identifiers and see if any source IP address appeared more than $\Psi\cdot \widehat{p}$ times; such a source IP is sending out to more than $\Psi$ destination IPs in expectation.

\section{Analysis}\label{sec:analysis}
In this section, we provide rigorous bounds on the accuracy of the algorithm.  
As a general note, we refer here to a \emph{packet stream} which can be distributed in any way between the measurement switches as long as each packet is measured at least once. 
All the results in this section are also applicable to flow sampling by replacing the notion to ``flow stream''. 
Specifically, we analyze the guarantee for estimating flow-sizes which, by simple reductions, also extend to Heavy Hitters (HH) and Hierarchical HH (HHH). For superspreaders the analysis is also applicable, although the condition regarding the minimum number of packets ($M$) is replaced by similar lower bound on the number of flows.
The entire section assumes that the hash functions are independent and are $\Omega(M)$-wise independent.  In practice, simpler hash functions often suffice~\cite{DBLP:journals/toc/ChungMV13}.

Our goal is to estimate flow sizes, with high probability, up to an additive error of $|\mathcal S|\cdot\epsilon$. This type of guarantee is standard in streaming algorithms and appears in~\cite{NetworkWideANCS,SpaceSavings,frequent4,ElasticSketch} and many others.
However, due to the nature of our algorithm, we cannot provide this guarantee immediately but rather require \emph{convergence time}. 
Recall that we first apply $h_1$ and map the packet into a slot that holds a single sample. Thus, it may take a while to achieve a large enough sample as multiple packets may be hashed to already full slots. Formally, \emph{hash collisions} in $h_1$ mean that some packets may not be sampled even if not all slots are full.

We mark by $\widetilde M(t)\in [0,\alpha \cdot M]$ the number of non-empty slots in our algorithm after seeing $t$ packets. 
We utilize the result of~\cite{NetworkWideANCS} that shows the accuracy guarantee one gets from analyzing a uniform sample of size $M$. By symmetry, we have that if $M(t)$ slots are filled, any subset of $M(t)$ packets has the same probability of appearing in the sample and therefore the sampling is uniform.
We say that our algorithm has \emph{converged} once $\widetilde M(t)\ge M$ and thus we provide the accuracy guarantee.
\begin{lemma}(\cite{NetworkWideANCS})\quad{}\label{lem:ANCS}
Let $T\subseteq\mathcal S$ be a random packet subset of size $M \ge \ceil{3\eps^{-2}\log_2(2/\delta)}$ . For a flow $x\in\mathcal U$, let $T_x$ be its  frequency in  $T$.
Then $\Pr\brackets{\bigg|f_x-T_x\cdot |\mathcal S|/M\bigg|\ge |\mathcal S| \epsilon}\le \delta$.
\end{lemma}

If $\alpha=1$, then the process of collecting the samples from the nonempty slots is known as the \emph{Coupon Collector} problem~\cite{blom2012problems}. In the Coupon Collector problem, a collector wishes to gather all $M$ coupons while getting a single coupon, uniformly at random, at each step. Since the time to collect the $i$'th distinct coupon is distributed geometrically with mean $M/(M-i)$, we have that the expected time to collect all coupons is $\sum_{i=1}^M M/(M-i) = M\ln M + O(M)$.
To derive a high-probability bound, observe that the probability that a given coupon is not collected after $r$ steps is $\parentheses{1-1/M}^r\le e^{-r/M}$. 
By using the union bound and setting $r=M\ln (M/\delta))$ we get that $\Pr\brackets{M(r)<M}\le M\cdot e^{-r/M}\le\delta$.
This analysis is directly applicable to our method for $\alpha=1$ as we uniformly hash every packet into one of the $M$ slots and the goal is to fill all slots. We can then choose $M$ to guarantee the desired result with probability $1-\delta/2$ and use the union bound to derive the standard $(\epsilon,\delta)$-guarantee.
We summarize this in the following theorem.
\begin{theorem}\label{thm:alpha=1}
For any $\epsilon,\delta>0$, let $M=\ceil{3\eps^{-2}\log_2(4/\delta)}$; our algorithm (with $\alpha=1$ and thus $M$ slots) guarantees approximating flow sizes up to an $(|\mathcal S|\epsilon)$-additive error, with probability $1-\delta$, given that the number of packets it processes is at least $M\cdot\ln(2M/\delta)$.
\end{theorem}

The above solution works well if the measurement is long enough with respect to the error parameters $\eps$ and $\delta$. However, this may prove to be too lengthy for accurate measurements. For example if $\epsilon=\delta=1\%$ we guarantee the \mbox{convergence of the algorithm} after about \mbox{$4.4$ million packets.} 

To shorten the convergence time, we explore the \emph{space to convergence time} tradeoff that $\alpha$ values larger than $1$ offer. Schematically, by increasing $\alpha$ we pay a constant factor in the amount of space required, but reduce the convergence time asymptotically, as we now show. We note that while the coupon collector analysis above is standard, to the best of our knowledge, the process described in this section is novel to our work. 
Particularly, we get that for any constant $\alpha>1$ the number of packets until convergence drops to $O(M)$, as summarized in the following theorem.
\begin{theorem}
Let $M\in\mathbb N^+,\alpha>1$ and denote $\beta=1+1/\ln\alpha+\ln(2/\delta)/(M\cdot\ln\alpha)$. When allocated with $\alpha \cdot M$ slots the algorithm fills at least $M$ of them after seeing $\beta\cdot M$ packets with probability at least $1-\delta/2$.
\end{theorem}
\begin{proof}
For a subset of indices $K\subseteq [\alpha \cdot M]$ of size $|K|=M-1$, define by $I_K$ an indicator for the event that all packets were mapped only into the indices of $K$. This event is bad as it means that the algorithm cannot produce an $M$-sized uniform packet sample and fails to provide the approximation guarantee. 
Observe that the probability of this event is $\Pr\brackets{I_K}=\parentheses{\frac{M-1}{\alpha \cdot M}}^{\beta\cdot M}$.
Since the number of such subsets $K$ is ${\alpha \cdot M \choose M-1}$, we can use the union bound to get that the probability that such a subset exists is at most 
$
\Pr\left[\exists K\subseteq [\alpha \cdot M]: \left|K\right|=M-1 \wedge I_K\right]\\ \le
    {\alpha \cdot M \choose M-1}\cdot \parentheses{\frac{M-1}{\alpha \cdot M}}^{\beta\cdot M}
    \le 
    (e\cdot\alpha)^M
    \cdot\parentheses{\frac{1}{\alpha}}^{\beta\cdot M} 
    = \delta/2,
    $
where the last inequality follows from the known binomial coefficient bound ${n\choose k}\le \parentheses{\frac{e\cdot n}{k}}^k$.
\end{proof}
In the following theorem, we once again use Lemma~\ref{lem:ANCS} for providing the error guarantee. 
To exemplify the reduction in convergence time, consider the above parameters ($\eps=\delta=1\%$ and $\alpha=2$ which means double space used). Our result implies guaranteed convergence after 630K packets, a reduction of over 85\%.
\begin{theorem}\label{thm:alpha>1}
For any $\epsilon,\delta>0$, let $M=\ceil{3\eps^{-2}\log_2(4/\delta)}$; AROMA (with $\alpha>1$ and $\alpha \cdot M$ slots) approximates flow sizes within an $(|\mathcal S|\epsilon)$-additive error, with probability $1-\delta$, after 
at least $M\cdot\parentheses{1+1/\ln\alpha+\ln(2/\delta)/(M\cdot\ln\alpha)}$ packets.
\end{theorem}

\newcommand{\SIGMETRICSFigureSize}{0.85\columnwidth}

\section{Evaluation}\label{sec:eval}
\textbf{Dataset:}
We used the CAIDA Anonymized Internet Trace 2018~\cite{CAIDA}.
 The trace contains internet packets collected from the  ``equinix-nyc'' high-speed monitor.
 For each packet, we use its 5-tuple (anonymized source-destination IP pair, port pair, and protocol) as its flow ID.
We summarize the number of distinct flows, for a given stream length, in Table~\ref{tbl:nrFlows}. 
\begin{table}[htb]
\centering
\resizebox{1\columnwidth}{!}{
  \begin{tabular}{|c|c|c|c|c|c|c|c|c|c|c|}
  \hline
  Length& $2^{16}$& $2^{17}$& $2^{18}$& $2^{19}$& $2^{20}$& $2^{21}$& $2^{22}$& $2^{23}$& $2^{24}$& $2^{25}$ \\\hline
  \#flows& $15$K& $26$K& $41$K& $66$K& $107$K& $183$K& $314$K& $550$K& $967$K& $1.69$M \\\hline
  
  \end{tabular}
}
\vspace{0.01cm}
\caption{The number of distinct $5$-tuples in the measurement as a function of the number of packets in the trace. }\label{tbl:nrFlows}
\ifdefined\arXiv
    \else
\cutSpaceAfterCaption
\fi
\end{table}
%

\textbf{Metrics:}
We consider the following performance metrics:
\begin{enumerate}[leftmargin=*]
\item Root Mean Square Error (RMSE): Measures the differences between predicted values of an estimator to actual values. Formally, for each flow $x$ the estimated frequency is $\widehat{f_{x}}$ and real frequency is $f_{x}$.  RMSE is calculated as: $\sqrt{\frac{1}{|\mathcal U|}\sum_{x\in\mathcal U}(\widehat{f_{x}}-f_{x})^{2}}$.

\item F1 Score: A quantity that combines precision (the correct fraction of reported flows), and recall (the fraction of true flows that were reported) into a single numerical value in the following manner: $F1= 2\cdot(precision\cdot recall)/(precision+recall)$.

\item Weighted Mean Relative Difference (WMRD): consider the set of flow sizes $\set{f_x\vert x\in \mathcal U}$ and let $z$ be the size of the largest flow. Denote by $F_i=\abs{\set{x\in \mathcal U\vert f_x=i}}$ denote the number of flows of size $i$. Let $\widehat{F_i}$ be the estimation produced by an algorithm for $F_i$. Define the sum of absolute errors to be $E=\sum_{i=1}^z\abs{F_i-\widehat{F_i}}$ and the sum of averages as $A=\sum_{i=1}^z (F_i+\widehat{F_i})/2$. The metric is then defined as $WMRD = E / A$. 
WMRD is always between $0$ and $2$ with a perfect match being $0$ and complete disagreement being $2$.

\end{enumerate}

\begin{figure*}[h]
    \ifdefined\arXiv
    \else
    \vspace*{-5mm}
    \fi
        \subfloat[Heavy Hitters]{\label{fig:heavyhitters_vsStreamLength}\includegraphics[width =0.5\columnwidth]
            {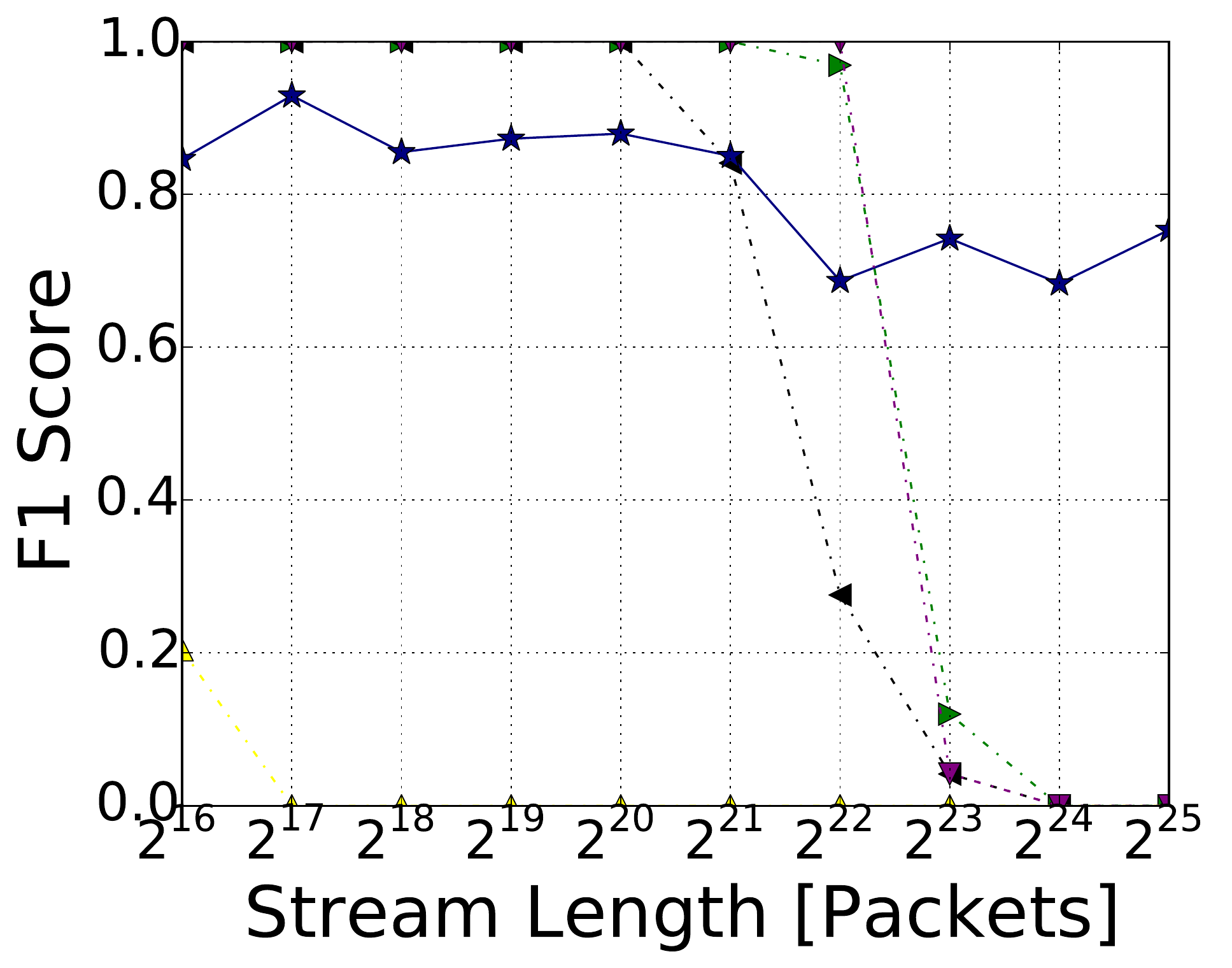}} 
            \hspace{.3cm}
        \subfloat[Hierarchical Heavy Hitters]{\label{fig:HHH_vsStreamLength}\includegraphics[width =0.5\columnwidth]
            {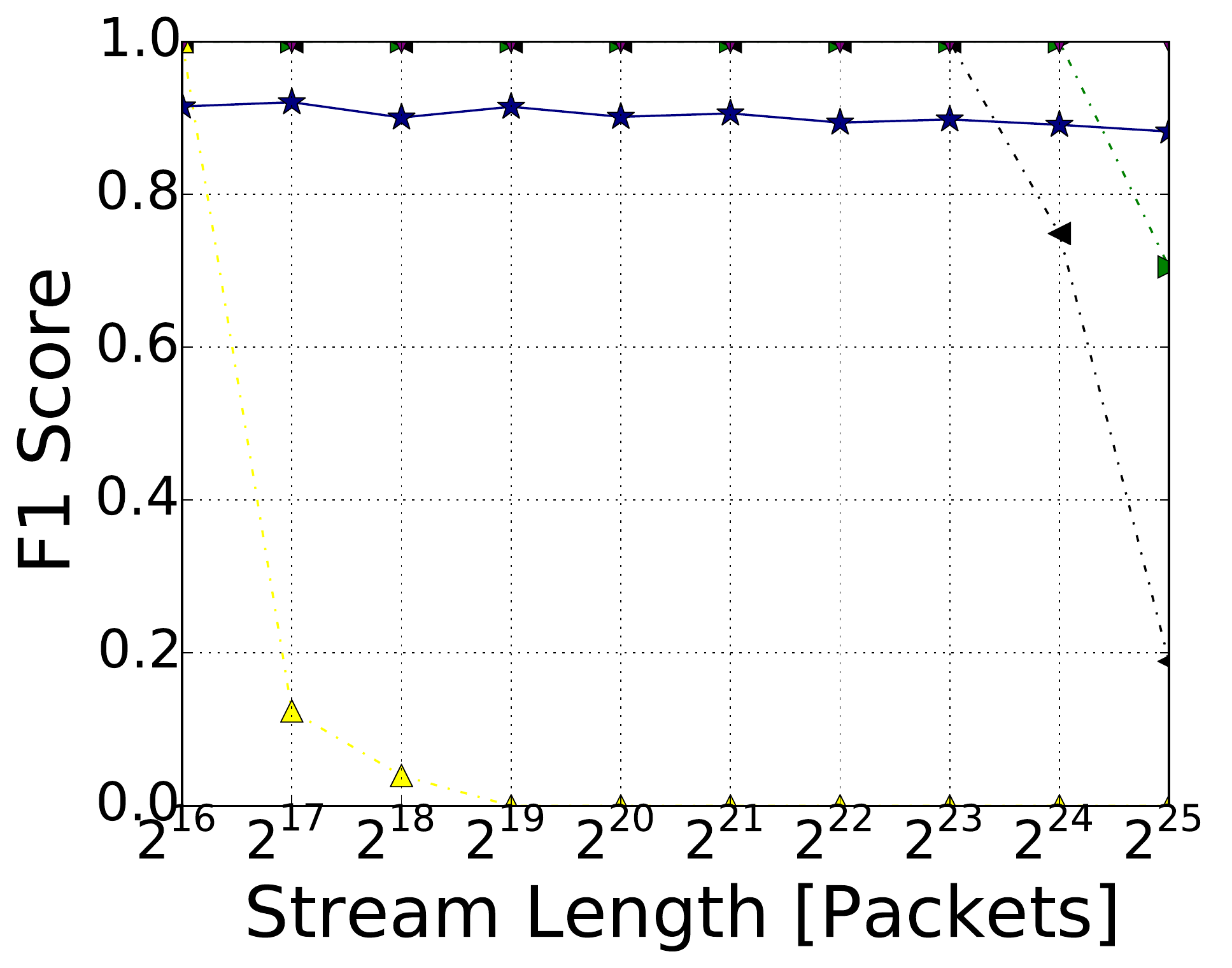}}
            \hspace{.3cm}
        \subfloat[Superspreaders]{\label{fig:superspreaders_vsStreamLength}\includegraphics[width =0.5\columnwidth]
            {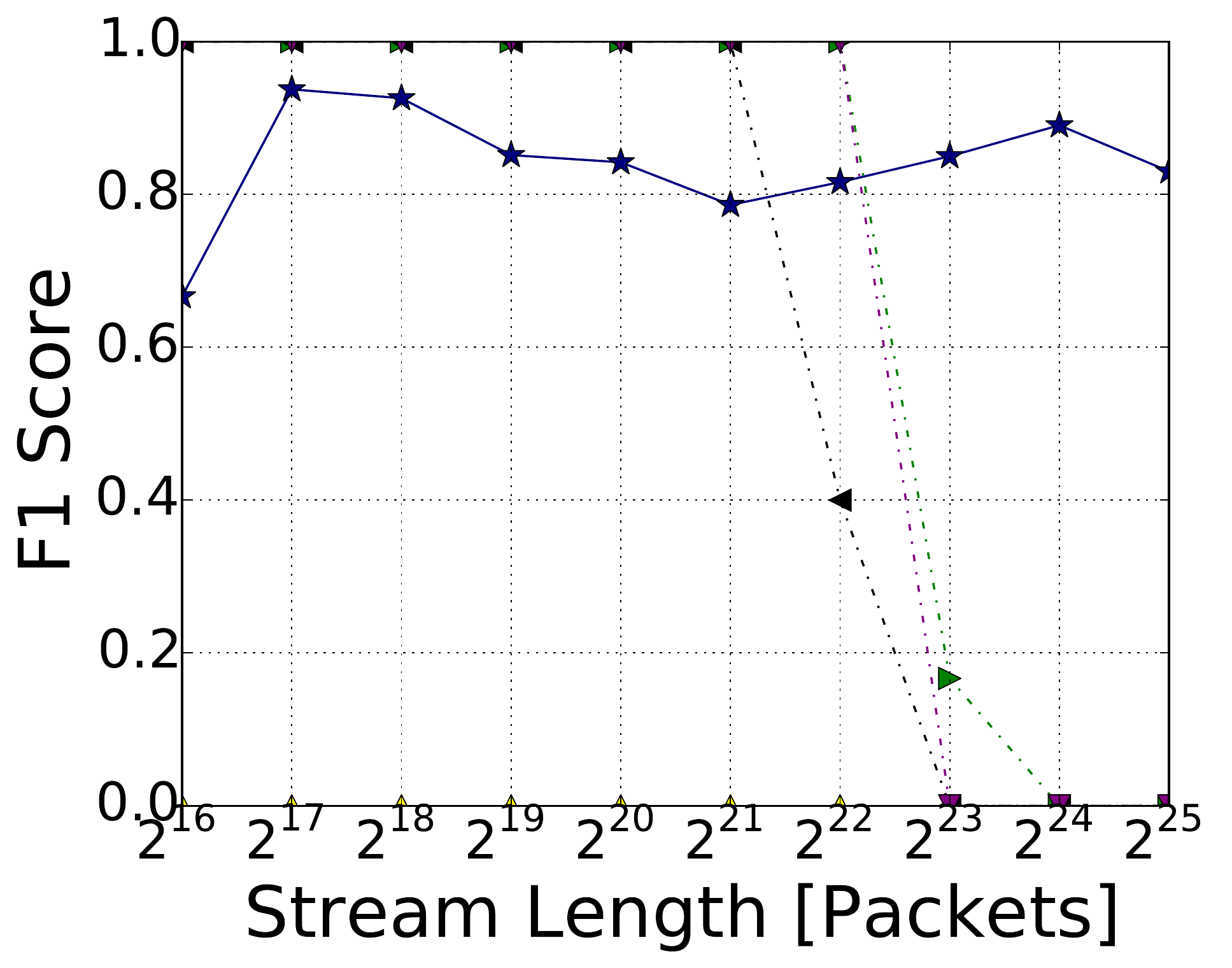}}
    \centering{
    \includegraphics[width=0.6\textwidth]{Figures/legend.png}}
    \caption{\label{fig:F1_vsStreamLength}{\mbox{F1 score (higher is better) on New York 2018, varying measurement tasks and per-switch space.}  }}
    \ifdefined\arXiv
    \else
    \cutSpaceAfterCaption
    \vspace{-.5cm} 
    \fi
\end{figure*}

\textbf{Evaluation Parameters:} 
In Figures~\ref{fig:Uniform}-\ref{fig:FE_FD} we used the first $2^{25}\approx 33.55$ million packets. The x-axis in these plots is the allocated \emph{per-switch} space.
We define a heavy hitter as a flow whose size is at least $0.1\%$ of the overall number of packets in the measurement. For the first $2^{25}$ packets of New York 2018 dataset this amounts to 35 heavy hitters. Similarly, we define a hierarchical heavy hitter as a source network that appear in more than $0.1\%$ of the overall traffic; this follows the HHH definition of~\cite{zhang2004online}.
We define a superspreader as a source IP that communicated with at least $\Psi=1000$ distinct destination IPs. For these parameters, we measured $54$ such~sources.
In Figures~\ref{fig:F1_vsStreamLength}-\ref{fig:FE_FD_vsStreamLength}, where the number of packets varies, we keep the $0.1\%$ threshold for HH and HHH and set the SS threshold such that there are $\approx50$ superspreaders for each~point.


\ifdefined\LARGEFIGURES
\begin{figure*}[bhtp]
\centering{\vspace{-0.2cm} 
\subfloat[Frequency Estimation]{\label{fig:frequencyestimationSample}\includegraphics[width =\SIGMETRICSFigureSize]
            {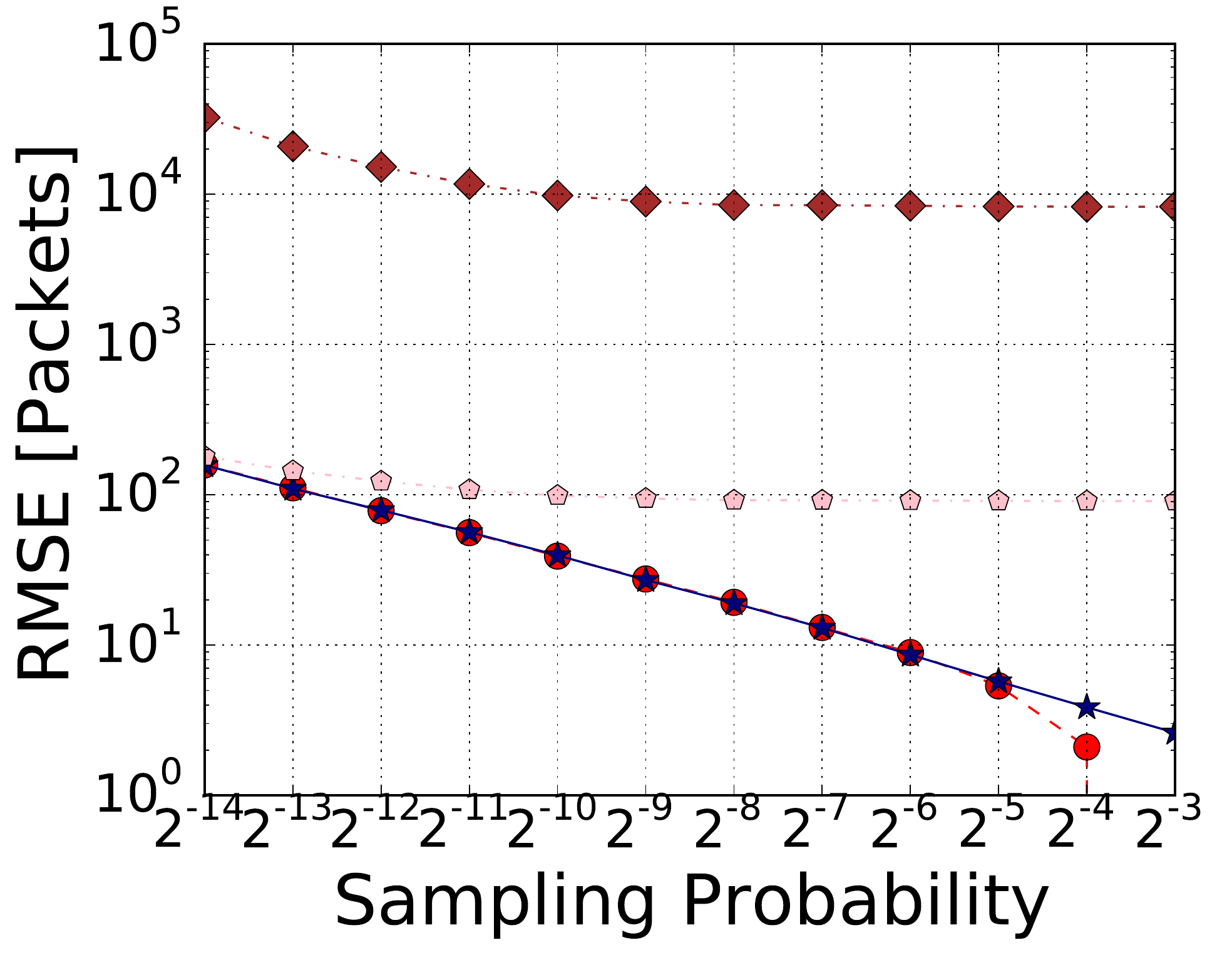}} 
        \subfloat[Heavy Hitters]{\label{fig:heavyHittersSample}\includegraphics[width =\SIGMETRICSFigureSize]
            {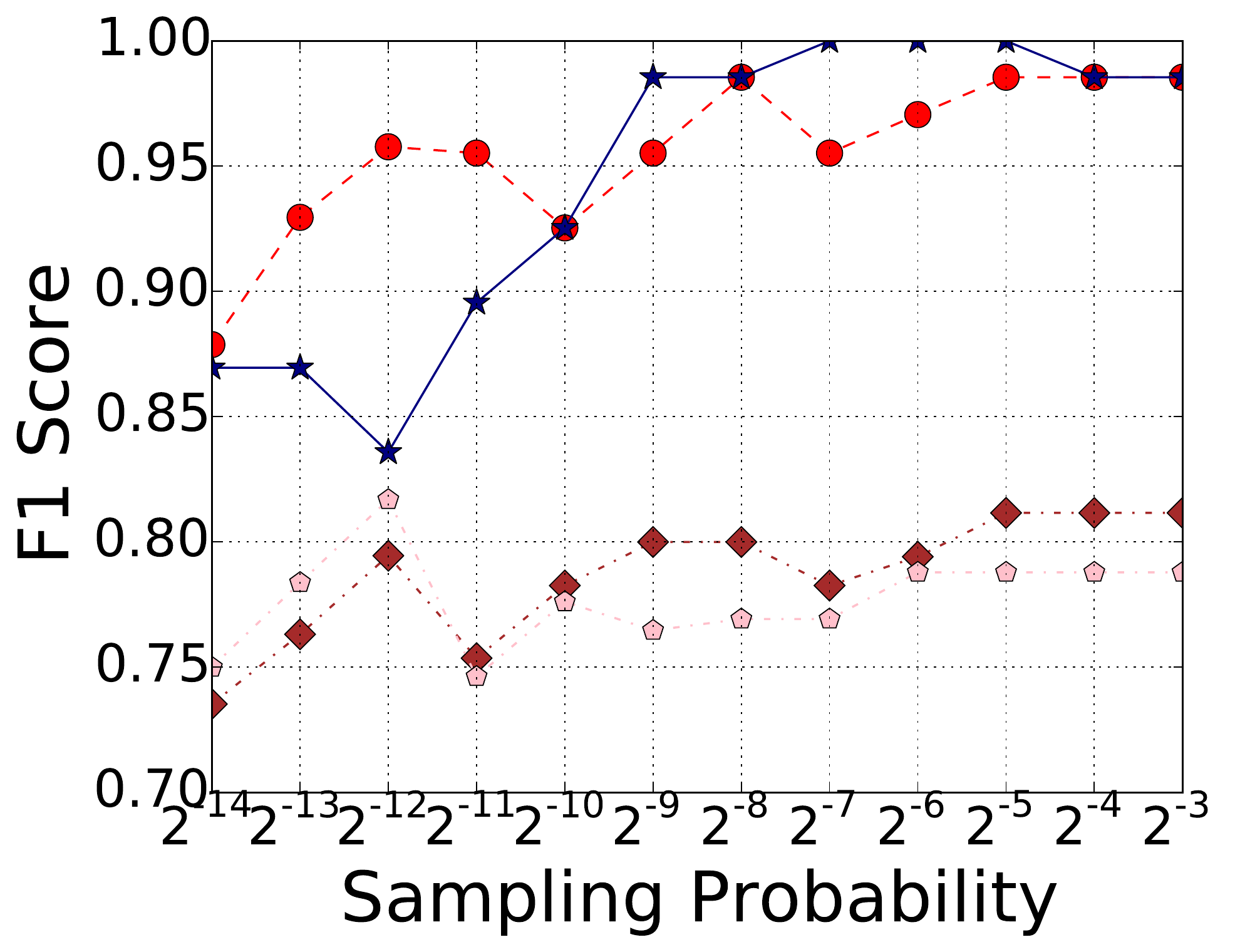}} \\
            \includegraphics[width=0.52\textwidth]{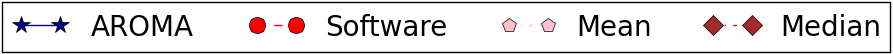}}
    \caption{\label{fig:Uniform}Root Mean Square Error for frequency estimation (lower is better), and F1 score for heavy hitters (higher is better), when comparing our method to (plain) random sampling, on 
    an Internet-like hop count distribution. 
    %
    }
    \cutSpaceAfterCaption
\end{figure*}
\else
\begin{figure}[t]
\ifdefined\arXiv
    \else
\vspace{-.5cm}
\fi
\centering{
        \subfloat[Frequency Estimation]{\label{fig:frequencyestimationSample}\includegraphics[width =0.49\columnwidth]
            {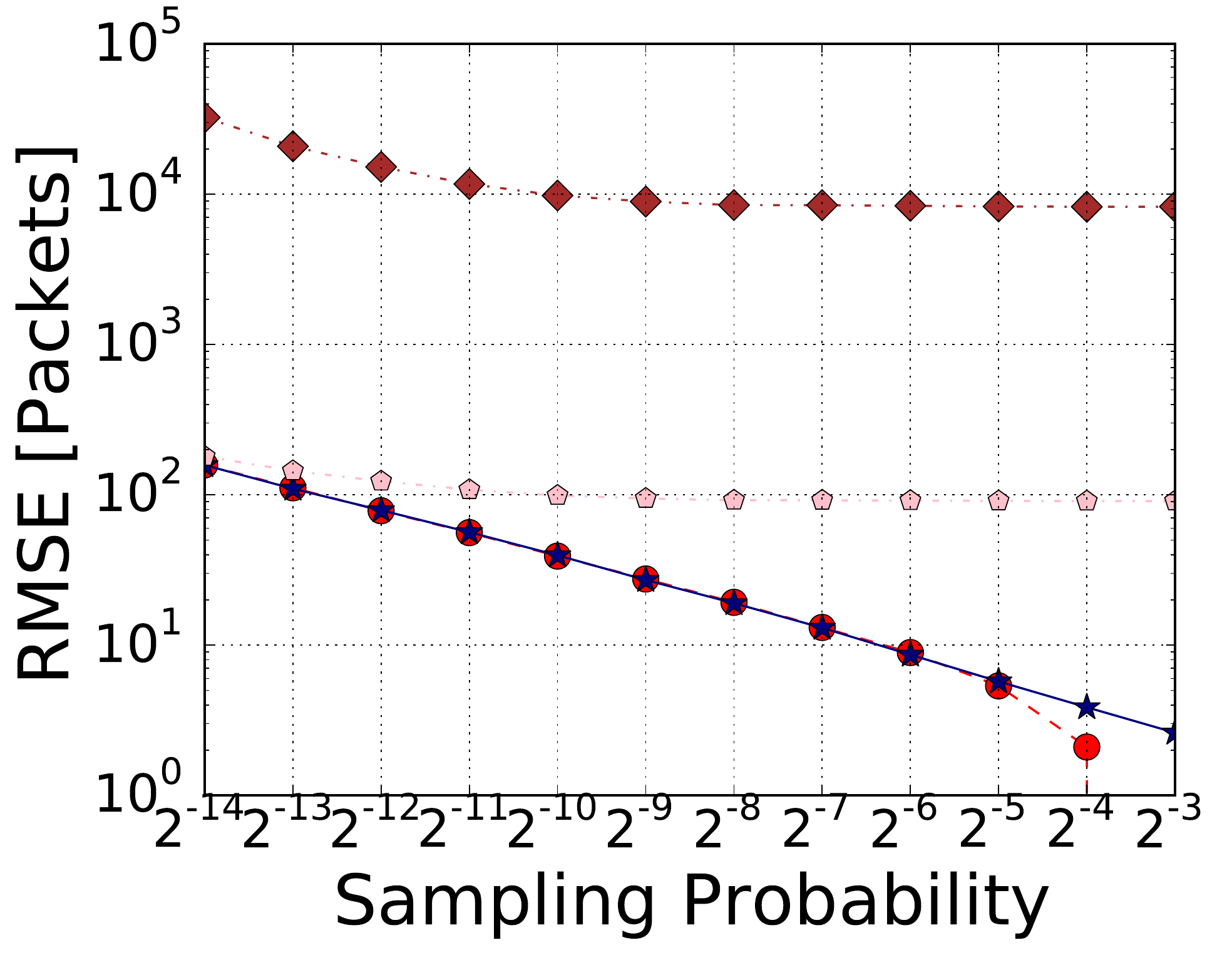}} 
        \subfloat[Heavy Hitters]{\label{fig:heavyHittersSample}\includegraphics[width =0.49\columnwidth]
            {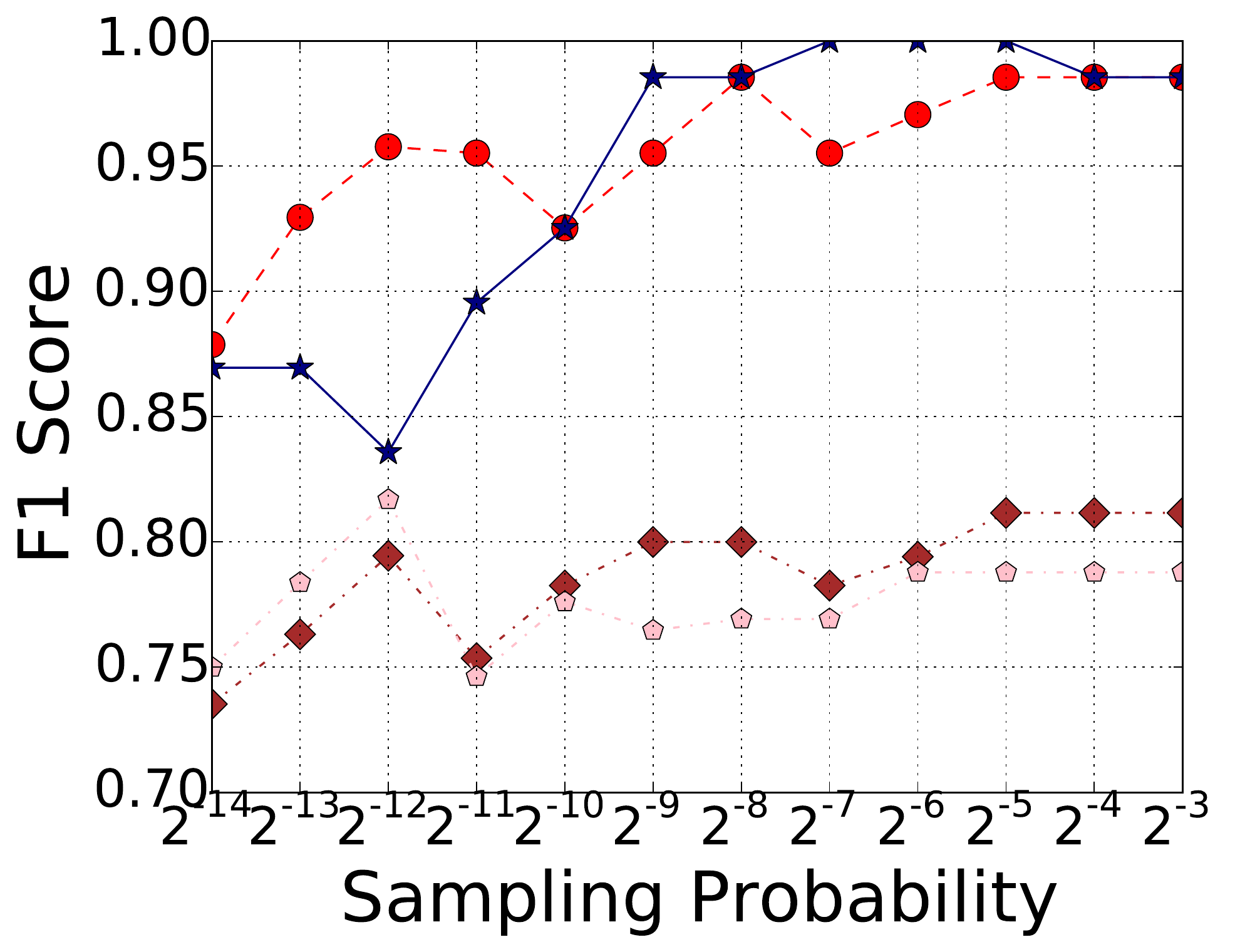}} \\
    \includegraphics[width=1.02\columnwidth]{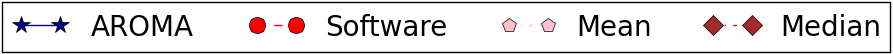}}
    \caption{\label{fig:Uniform}Root Mean Square Error for frequency estimation (lower is better), and F1 score for heavy hitters (higher is better), when comparing our method to (plain) random sampling, on 
    an Internet-like hop count distribution. 
    %
    }
    \ifdefined\arXiv
    \else
    \cutSpaceAfterCaption
    \vspace{-.4cm} 
    \fi
\end{figure}
\fi

\begin{figure*}[t!] 
\ifdefined\arXiv
    \else
\vspace{-.5cm}
\fi
    \begin{tabular}{ccc}
        \subfloat[Heavy Hitters]{\label{fig:heavyhitters}\includegraphics[width =0.5\columnwidth]
            {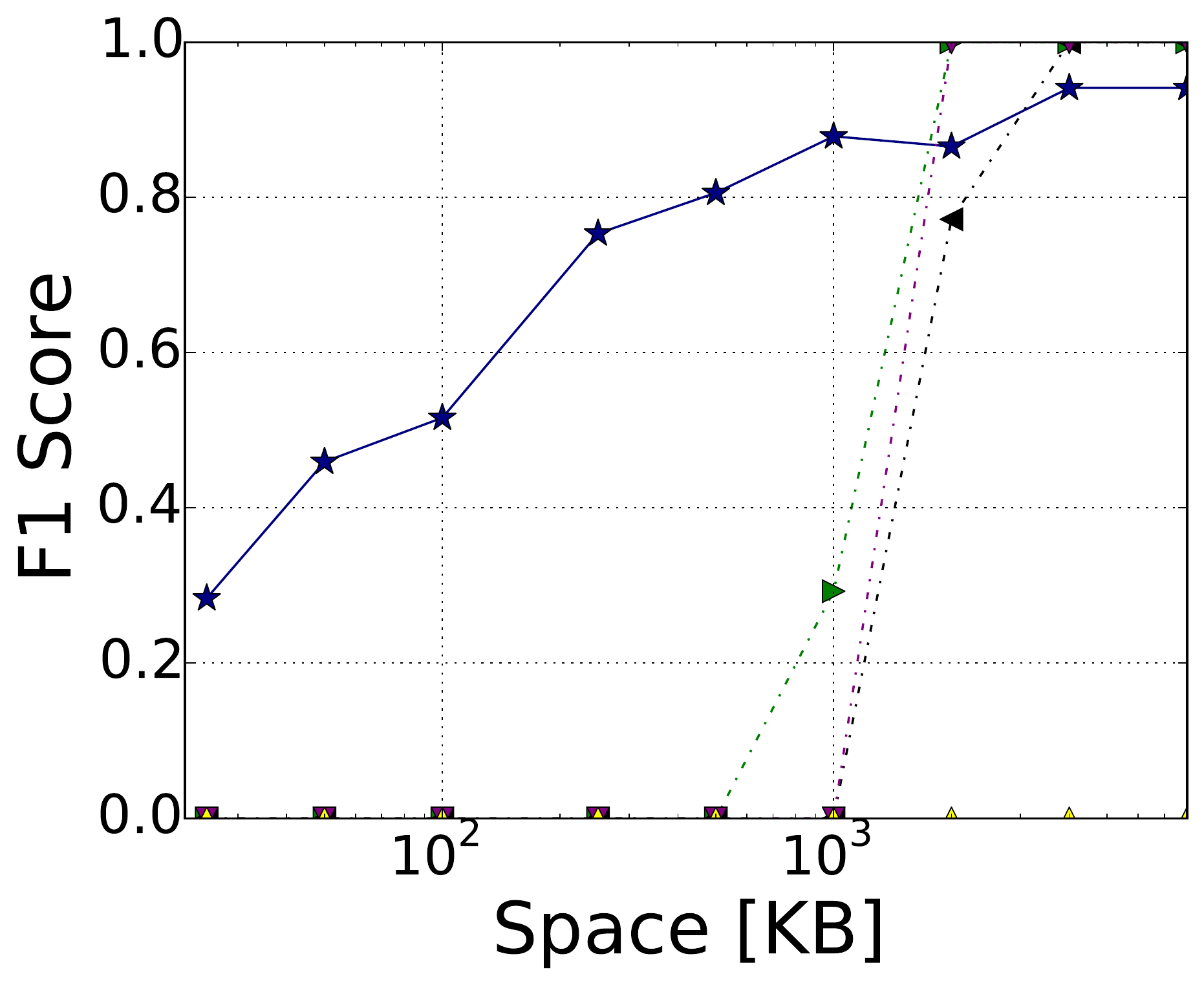}} &\hspace{.3cm}
        \subfloat[Hierarchical Heavy Hitters]{\label{fig:HHH}\includegraphics[width =0.5\columnwidth]
            {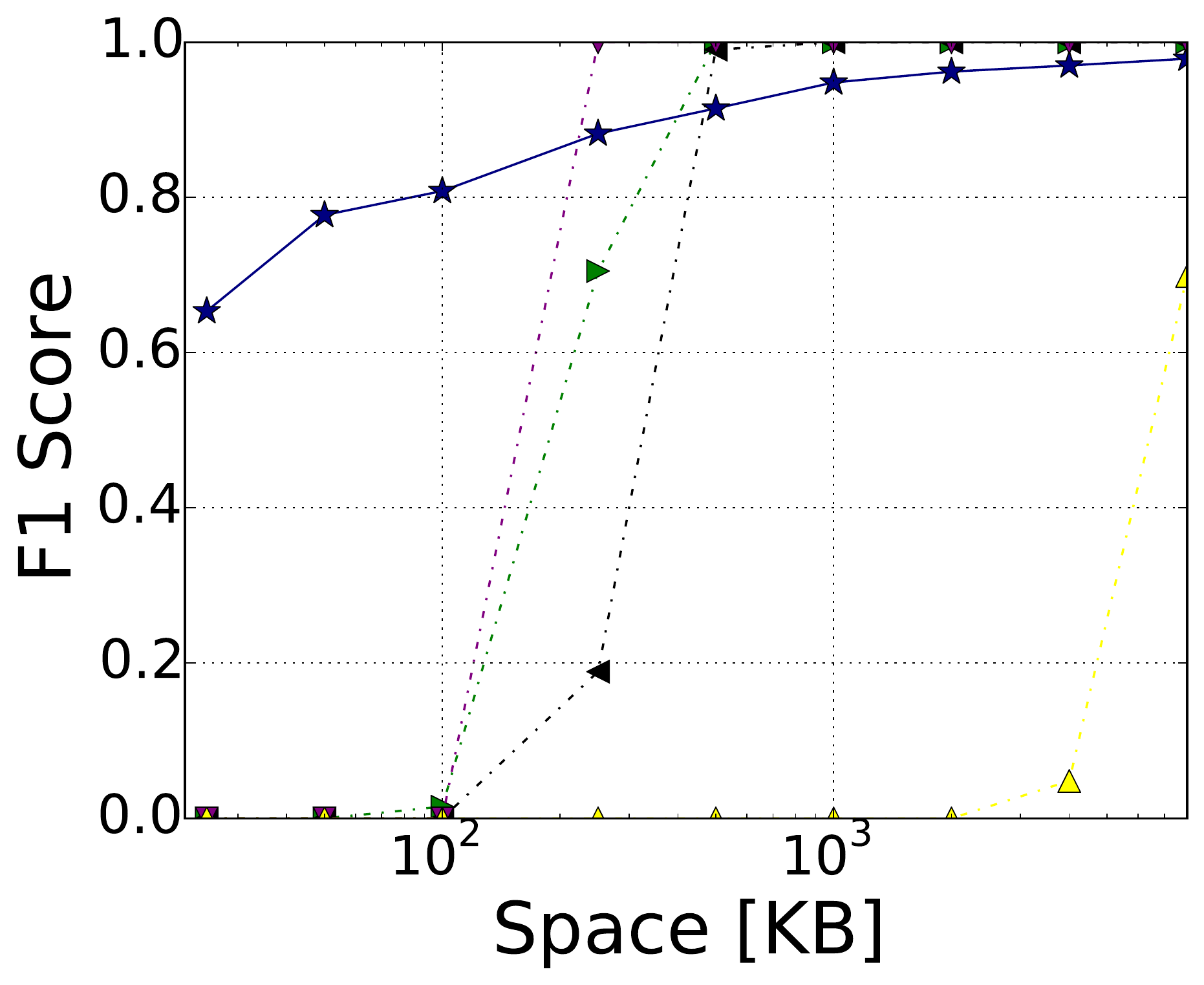}} &\hspace{.3cm}
        \subfloat[Superspreaders]{\label{fig:superspreaders}\includegraphics[width =0.5\columnwidth]
            {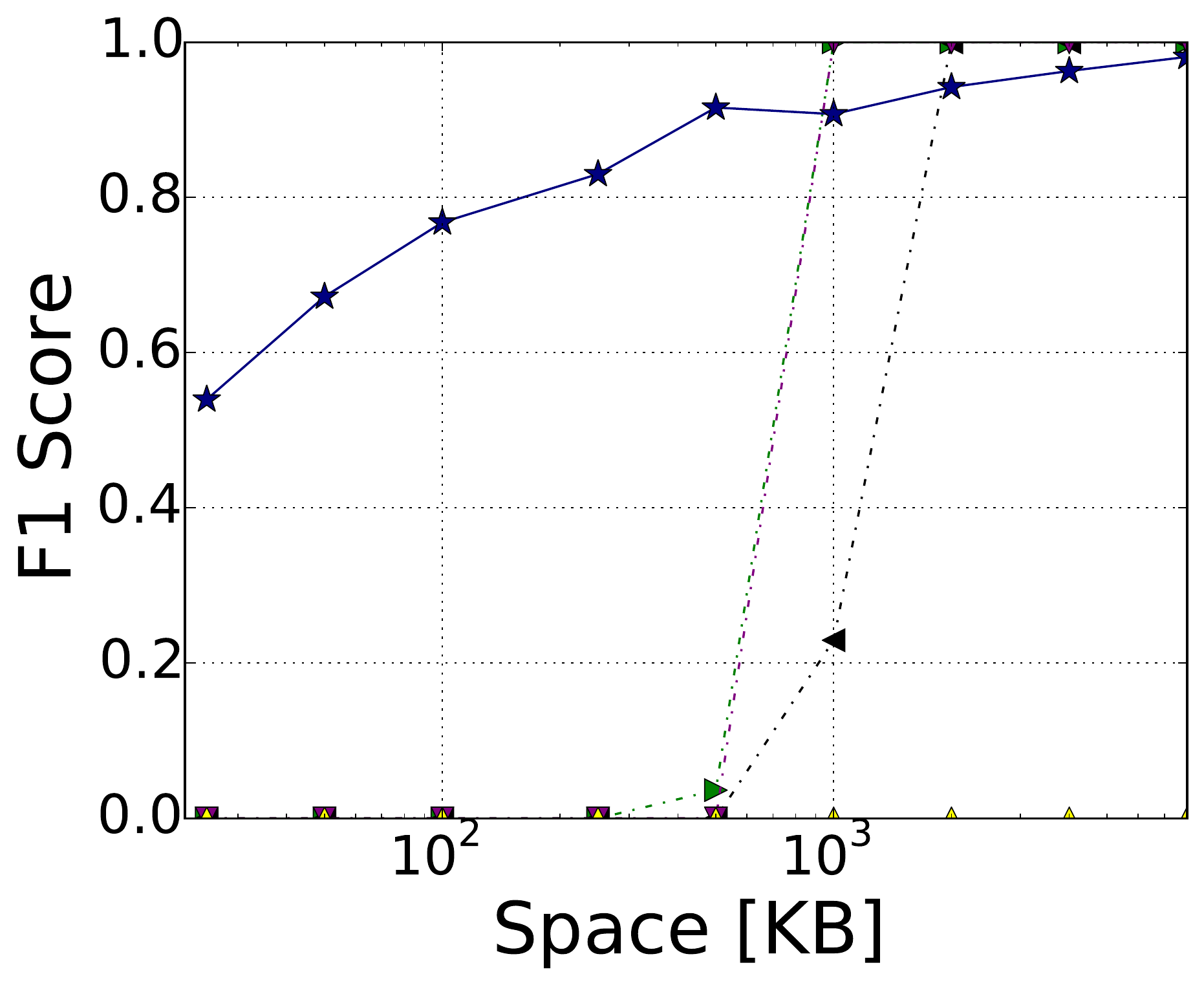}}
    \end{tabular}
    \ifdefined\arXiv
    \else
    \vspace{-.05cm} 
    \fi
    \centering{
    \includegraphics[width=\columnwidth]{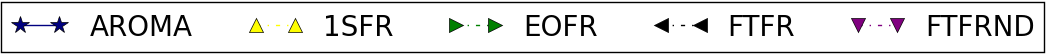}}
    \caption{\label{fig:F1}\mbox{F1 scores (higher is better) on New York 2018, for various measurement tasks, and per-switch space.}  }
    \ifdefined\arXiv
    \else
    \cutSpaceAfterCaption
    \vspace*{-.5cm}
    \fi
\end{figure*}

\ifdefined\LARGEFIGURES
\begin{figure*}[htbp]
        \subfloat[Frequency Estimation]{\label{fig:frequencyestimation}\includegraphics[width =\SIGMETRICSFigureSize]
            {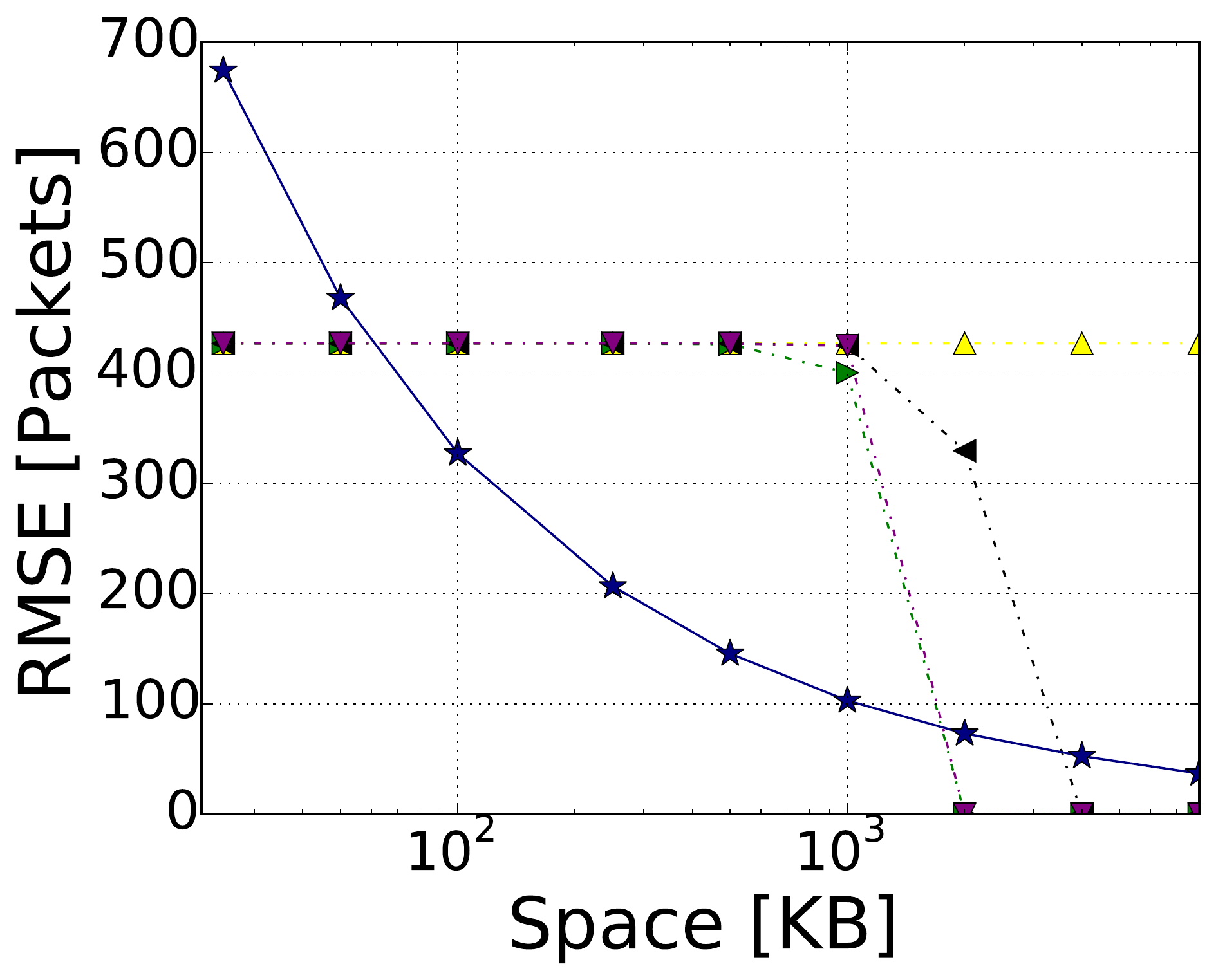}} 
        \subfloat[Flow Size Distribution Estimation]{\label{fig:flowsize}\includegraphics[width =\SIGMETRICSFigureSize]
            {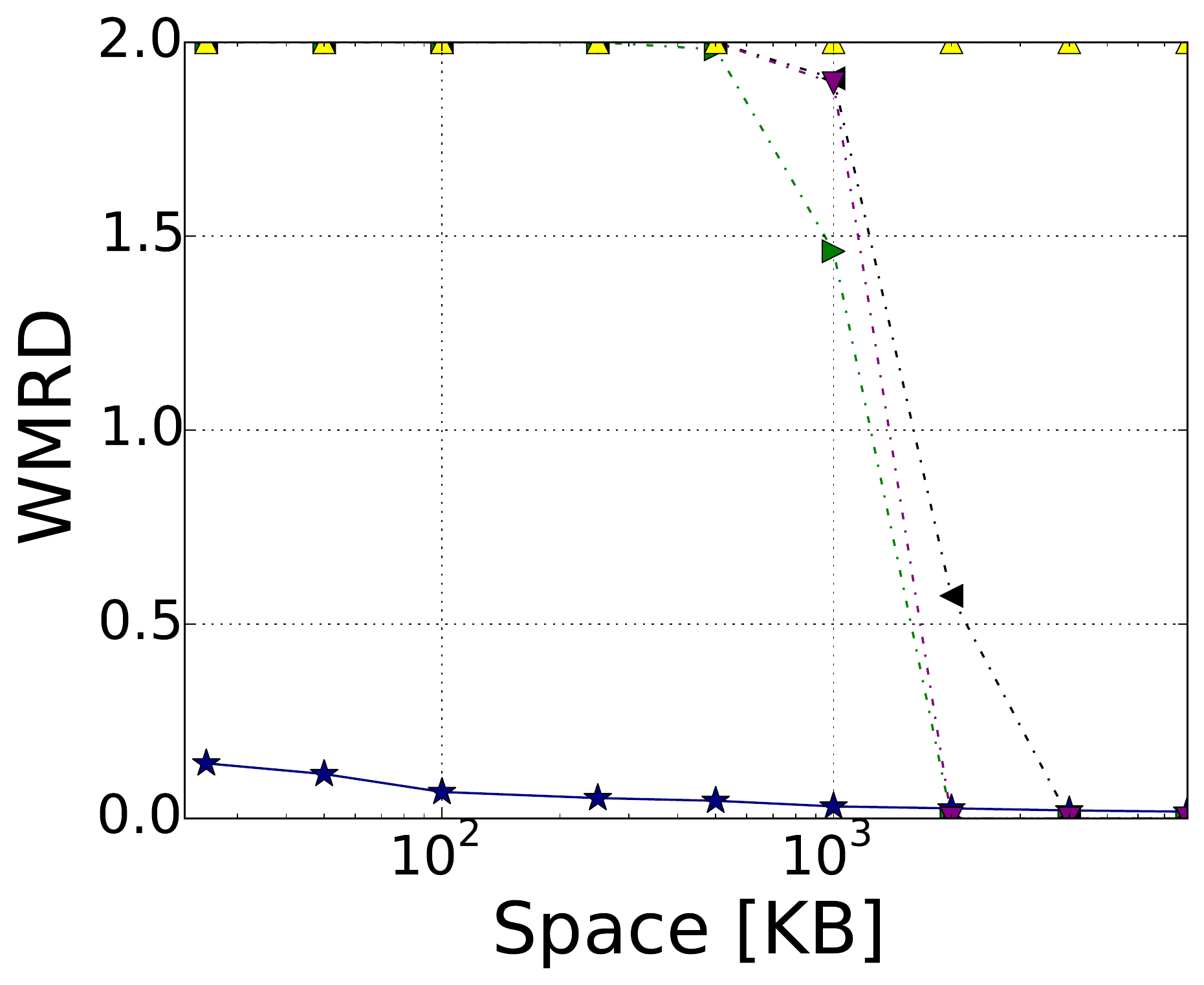}} 
    \centering{
    \includegraphics[width=0.6\textwidth]{Figures/legend.png}}
    \caption{\label{fig:FE_FD}Root Mean Square Error, and Weighted Mean Relative Difference  (lower is better), for the frequency estimation, and flow size distribution estimation tasks on the New York 2018 dataset. }
    \cutSpaceAfterCaption
\end{figure*}
\else
\begin{figure}[htbp]
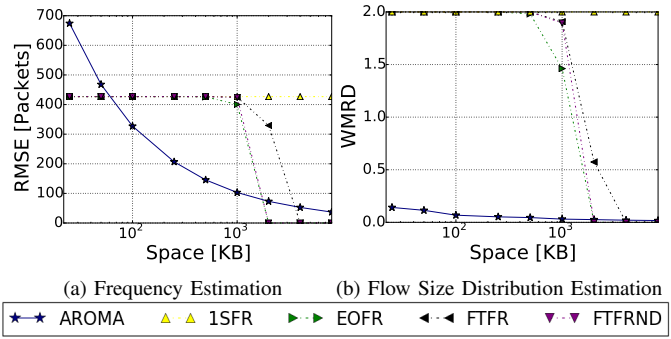

    \vspace{-.5cm}
        \subfloat[Frequency Estimation]{\label{fig:frequencyestimation}\includegraphics[width =0.49\columnwidth]
            {Figures/FreqEstRMSE_NY2018_FS}} 
        \subfloat[Flow Size Distribution Estimation]{\label{fig:flowsize}\includegraphics[width =0.49\columnwidth]
            {Figures/SizeDistributionWMRD_NY2018_FSD}} 
    \centering{
    \includegraphics[width=\columnwidth]{Figures/legend.png}}
    \caption{\label{fig:FE_FD}Root Mean Square Error, and Weighted Mean Relative Difference  (lower is better), for various tasks on the New York 2018 dataset. }
    \ifdefined\arXiv
    \else
    \cutSpaceAfterCaption
    \vspace{-.4cm}
    \fi
\end{figure}
\fi

\textbf{Comparison with uniform sampling and software solutions:}
We start comparing AROMA's accuracy to naive uniform sampling. Recall that in such sampling, packets get an opportunity to be sampled for each measurement switch they visit, which biases the controller's sample.  Additionally, we compare AROMA to the BEFMR18 software routing oblivious algorithm of~\cite{NetworkWideANCS}. 
We used the Internet's hop count distribution by~\cite{HopCount,HopCount2}, and assumed a measurement switch at each hop which improves reliability.
Specifically, this model assumes that the probability for $k$-hops (for a given flow) is:\\
{\small $
\Pr[k\mbox{ hops}] = \frac{1+o(1)}{N}\sum_{m=0}^k c_{m+1}\frac{(\ln N)^{k-m}}{(k-m)!},
$}
where $c_{i}$ is the $i$'th Taylor coefficient of the reciprocal of the Gamma function $1/\Gamma(z)$~\cite[Table~6.1.36]{abramowitz1965handbook}.
The work of~\cite{HopCount2} models the actual hop-count distribution of the Internet as the distribution for $N=98400$, which gives a median hop count of $12$.

We deploy measurement switches on each hop and normalize the frequency at the controller by either the mean or the median hop count, of the hop count distribution. 
Figure~\ref{fig:Uniform} shows the results of this evaluation, where Median (Mean) is the uniform sampling normalized by the median (mean) value, AROMA is our algorithm and Software refers to~\cite{NetworkWideANCS}. Figure~\ref{fig:frequencyestimationSample} shows the results for estimating per flow frequency. The mean normalization provides better accuracy for (plain) random sampling. Our method and Software are almost identical and are considerably more accurate for a wide range of sampling probabilities.  Figure~\ref{fig:heavyHittersSample} shows the F1 score for heavy hitters; our method and the Software method offer higher F1 values than uniform sampling. Note that for this application, it is unclear which normalization (mean or median) is superior. Intuitively, uniform sampling suffers from flows whose hop-count significantly differs from
the mean or median and thus are grossly \mbox{underestimated or overestimated.}
%

\textbf{Comparison with existing hardware solutions:}
Figure~\ref{fig:F1} shows an evaluation of AROMA's performance when compared to existing works, for various measurement tasks. 
We also compared with FlowRadar~\cite{FlowRadar} in various configurations. Throughout the evaluation, we consider FlowRadar's estimation of the size of a flow which it failed to decode as~zero. For the FatTree topology we assume that the flows are distributed uniformly, the easiest setting for FlowRadar.
\begin{itemize}[leftmargin=*]
    \item \texttt{\textbf{1SFR}} - FlowRadar where all packets go through a single switch.
    \item  \texttt{\textbf{FTFR}} - FlowRadar deployed on all switches of a k=8 fat tree with FlowDecode (the faster decoding procedure).
    \item \texttt{\textbf{FTFRND}} - FlowRadar deployed on all switches of a k=8 fat tree with NetDecode (the more accurate but slower decoding~procedure).  
    \item \texttt{\textbf{EOFR}} - FlowRadar deployed on all edge switches of a k=8 fat tree with packets only measured once (on the first edge switch they visit).
    \vspace{.1cm}
\end{itemize}

Figure~\ref{fig:heavyhitters} shows the F1 metric for heavy hitter measurement (higher F1 values are better).  
FlowRadar (in all scenarios) fails to provide any meaningful information until circa 1 MB of space, and from that point on, it rapidly improves with more space until it provides an exact measurement (F1=1) which is better than our approach. 
  We conclude that FlowRadar has a strict minimum memory size requirement while AROMA performs better under a tight memory budget, which is expected in today's switches, with only a few MBs of data-plane memory  shared among various measurement tasks.

Figure~\ref{fig:HHH} show results for the hierarchical heavy hitters' task, and Figure~\ref{fig:superspreaders} for superspreader measurements, as can be observed the qualitative behavior is the same as in the heavy hitter case. 
AROMA can operate and provide accurate measurements while FlowRadar fails unless it is allocated with enough space.
Thus, for these tasks, our algorithms are superior when given a small amount of space, and inferior when there is enough space to run \mbox{FlowRadar efficiently.}
Recalling Table~\ref{tbl:nrFlows}, we observe that the actual number of flows increases with the measurement length. Thus, we expect our method to be more reliable in long measurements, especially as traffic anomalies such as port scan attacks can increase the number of flows in the measurement.

Figure~\ref{fig:FE_FD} shows results for our packet sampling algorithm and the frequency estimation problem. As well as, for our flow sampling algorithm, and the flow size distribution estimation problem.
In Figure~\ref{fig:frequencyestimation}, we can see that our approach
continuously improves given more space. In contrast, the various FlowRadar configurations are very inaccurate until there is enough memory, and then they have no error at all. Still, AROMA outperforms \mbox{FlowRadar in many configurations.}

In Figure~\ref{fig:flowsize}, we see results for the flow size distribution estimation task. 
In the FlowRadar configurations, we again see the ``cliff'' where the algorithms do not work until there is enough memory allocated. Notice that the required memory for them to work is several megabytes, whereas our algorithm is accurate even with a few kilobytes.

Next, we allocate 250KB for each switch and monitor the accuracy throughout the trace. Figure~\ref{fig:F1_vsStreamLength} shows the F1 score for different applications and varying the stream length. Initially, FlowRadar configurations achieve accurate measurement (F1 score of 1). Then, as the measurement prolongs, we encounter more flows, and FlowRadar configurations begin to fail. Once we reach 32 million packets, all the FlowRadar configurations become ineffective. In contrast, our sampling-based approach 
is 
relatively accurate throughout the measurement. We conclude that AROMA is superior when there is insufficient \mbox{memory space for an accurate measurement.}
While accurate measurements are desired, it is unclear how much memory FlowRadar would need to succeed, whereas in AROMA we always yield a relatively accurate outcome.

Figure~\ref{fig:FE_FD_vsStreamLength} shows results for the frequency estimation, and flow size distribution estimation tasks (lower is better). In Figure~\ref{fig:frequencyestimationSample_vsStreamLength}, we see that the accuracy of our method gracefully degrades throughout the measurement. FlowRadar degrades accuracy less gracefully as the measurement prolongs. When the measurement is long enough, our approach is more accurate than all FlowRadar configurations. Figure~\ref{fig:flowsize_vsStreamLength} shows the flow size distribution estimation accuracy throughout the measurement. The results are qualitatively similar, but our method does much better than \mbox{FlowRadar configurations.}

\textbf{Performance breakdown:}
For the HH, HHH, and SS tasks, we used F1 as a single-metric for evaluating the performance of algorithms. However, the actual precision and recall performance of the algorithm is not the same. Our algorithms provide near-perfect precision and recall, while FlowRadar gives perfect precision but a poorer recall. The reason is that FlowRadar provides the exact sizes of the flows it decodes and thus know if one is a heavy hitter. We show the precision and recall performance in Figure~\ref{fig:precisionRecall}.

\ifdefined\LARGEFIGURES
\begin{figure*}[tbhp]
        \subfloat[Frequency Estimation]{\label{fig:frequencyestimationSample_vsStreamLength}\includegraphics[width =\SIGMETRICSFigureSize]
            {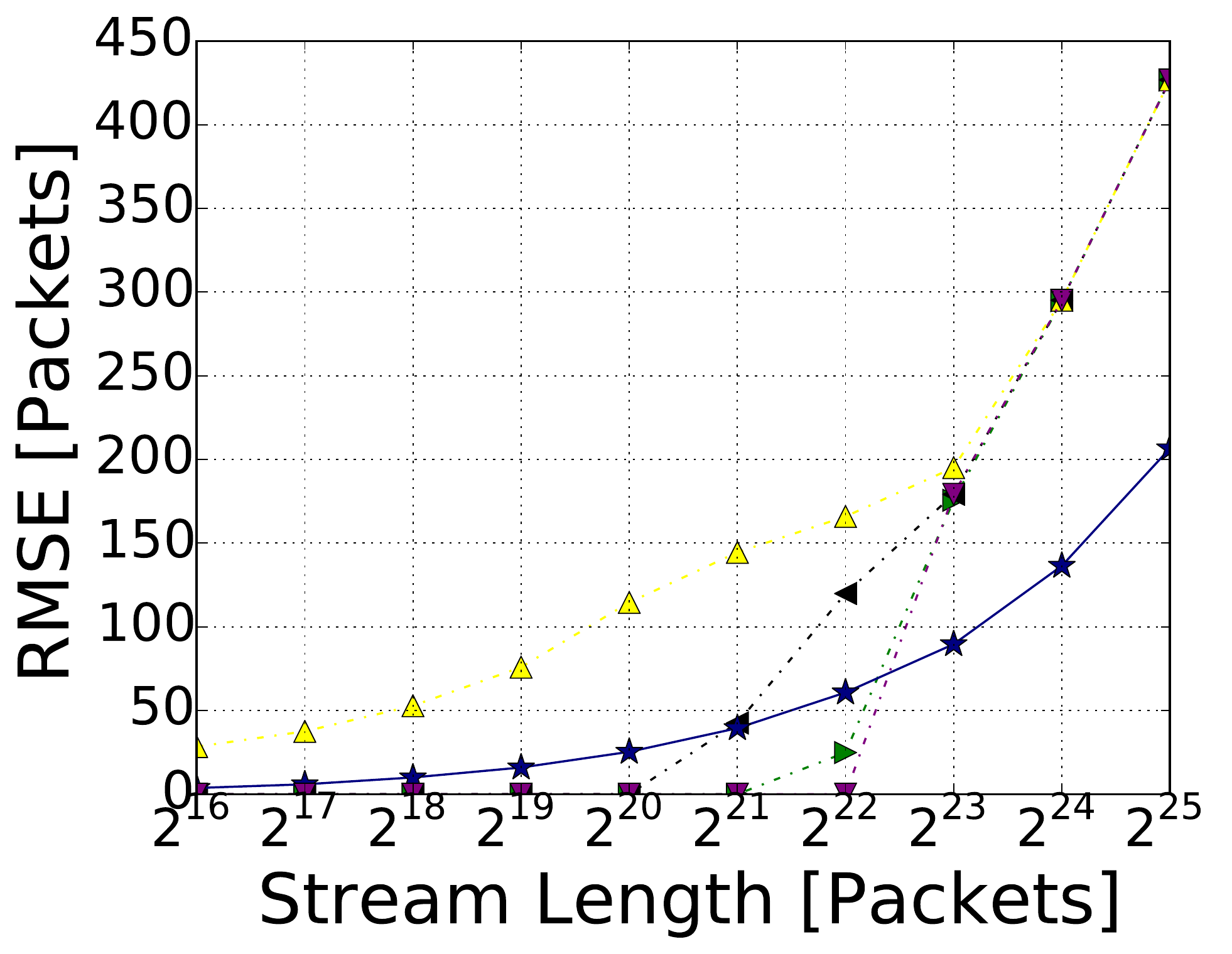}} 
        \subfloat[Flow Size Distribution Estimation]{\label{fig:flowsize_vsStreamLength}\includegraphics[width =\SIGMETRICSFigureSize]
            {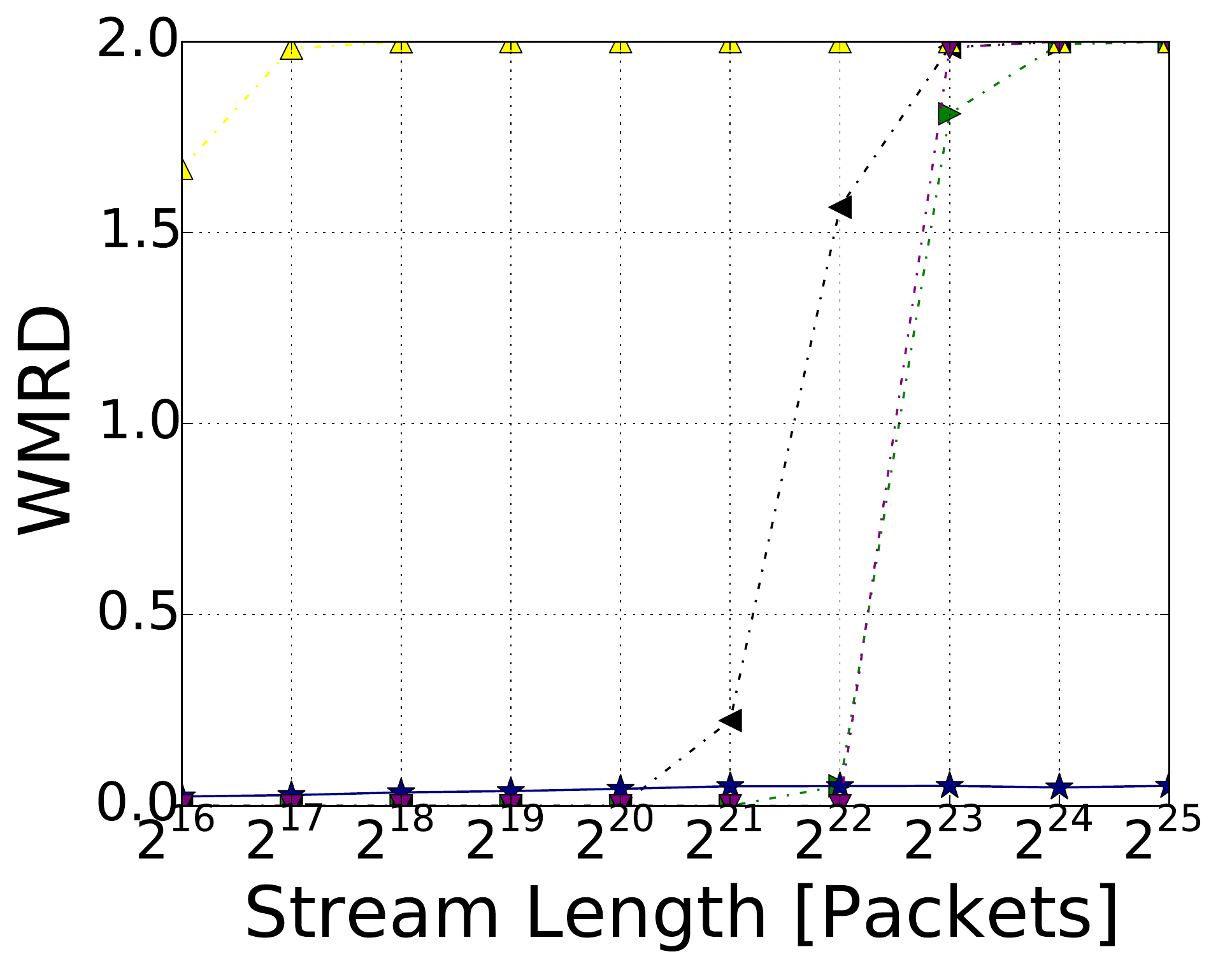}} 
    \centering{
    \includegraphics[width=0.6\textwidth]{Figures/legend.png}}
    \caption{\label{fig:FE_FD_vsStreamLength}Root Mean Square Error and Weighted Mean Relative Difference  (lower is better)  on the New York 2018 dataset, varying the measurement's length.
    }
    \ifdefined\arXiv
    \else
    \cutSpaceAfterCaption
    \vspace{-.4cm}
    \fi
\end{figure*}
\else
\begin{figure}[t]
    \ifdefined\arXiv
    \else
    \vspace*{-.5cm}
    \fi
        \subfloat[Frequency Estimation]{\label{fig:frequencyestimationSample_vsStreamLength}\includegraphics[width =0.49\columnwidth]
            {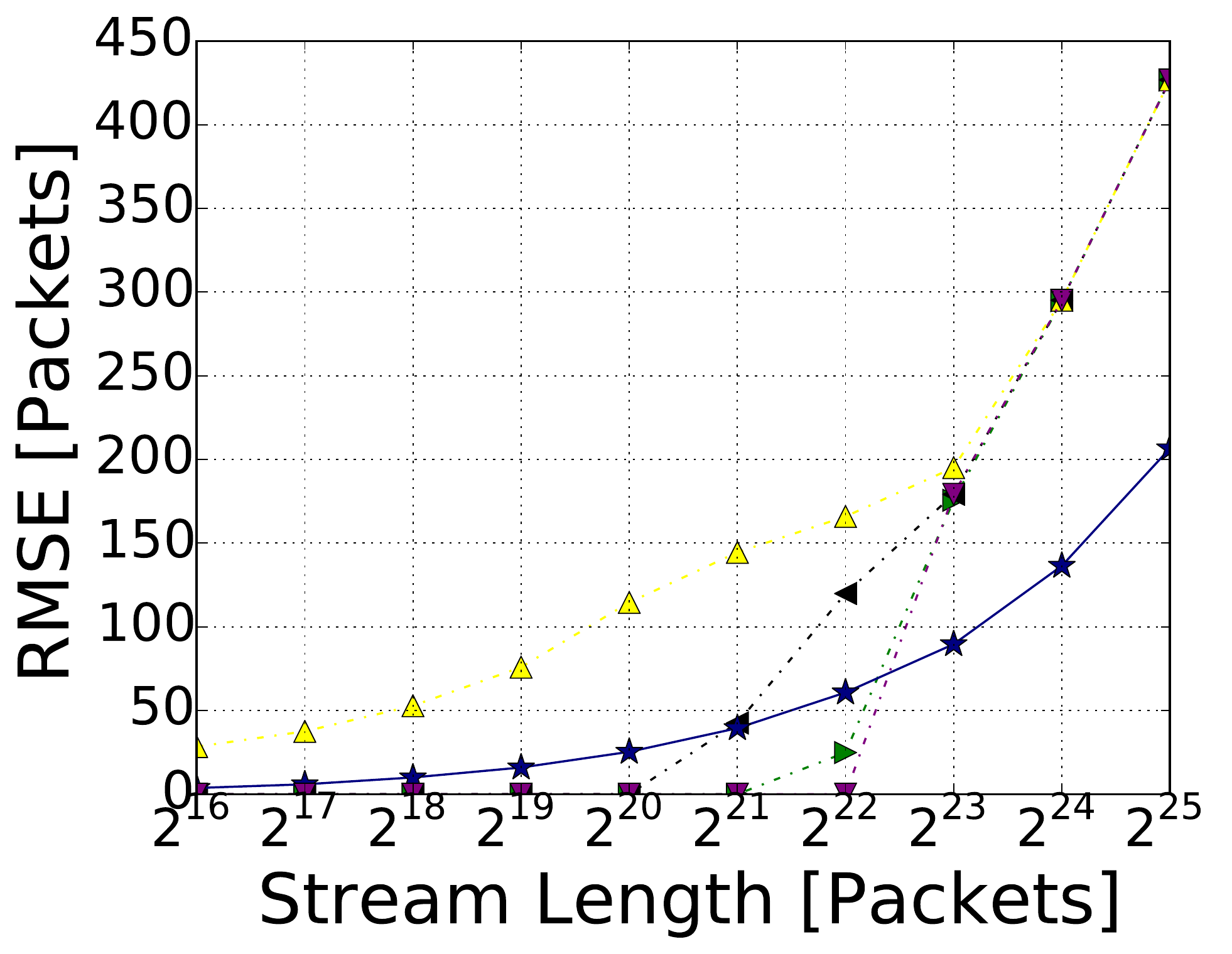}} 
        \subfloat[Flow Size Distribution Estimation]{\label{fig:flowsize_vsStreamLength}\includegraphics[width =0.49\columnwidth]
            {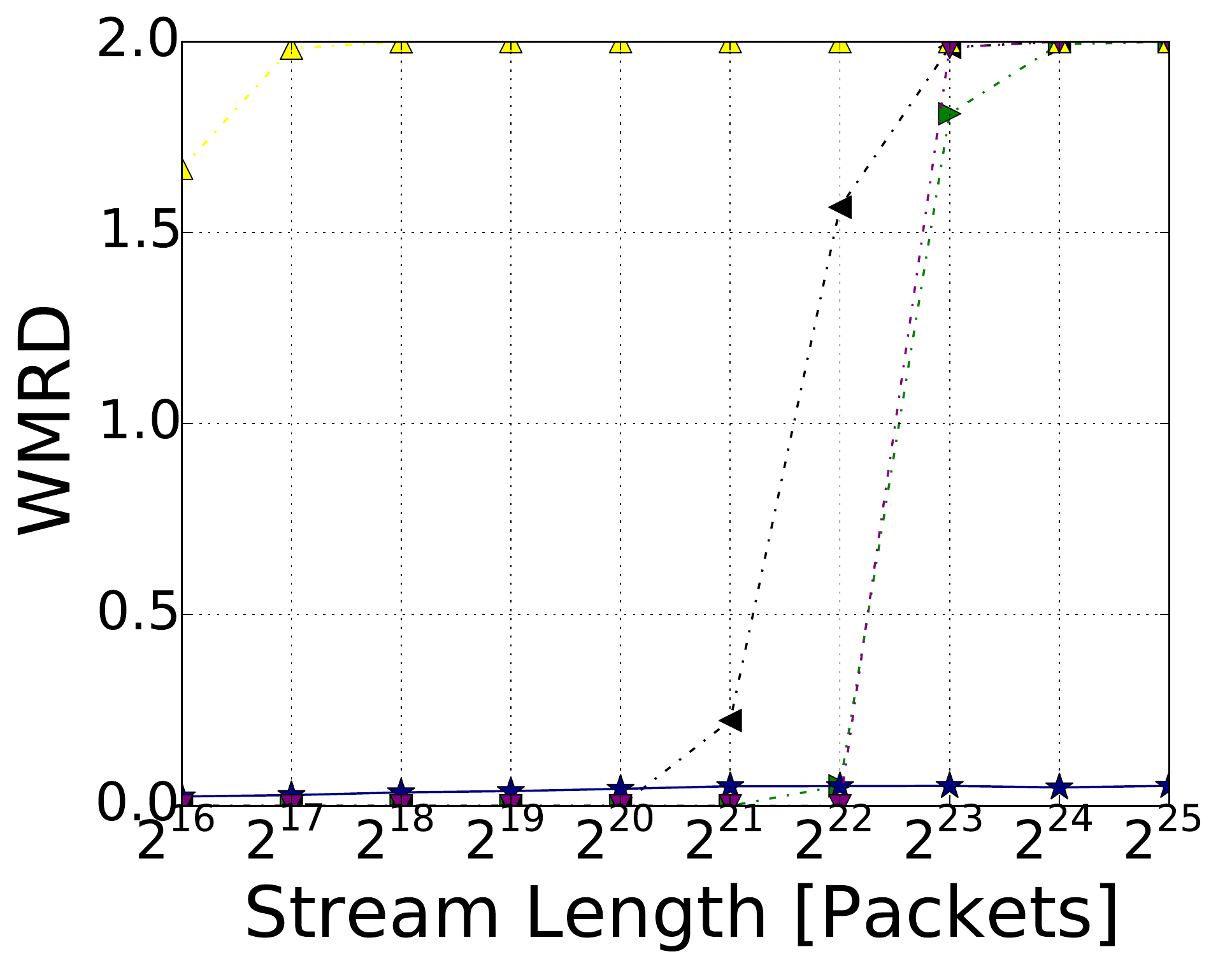}} 
    \centering{
    \includegraphics[width=\columnwidth]{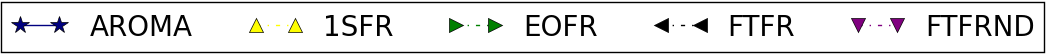}}
    \caption{\label{fig:FE_FD_vsStreamLength}Root Mean Square Error and Weighted Mean Relative Difference  (lower is better)  on the New York 2018 dataset, varying the measurement's length.
    }
    \ifdefined\arXiv
    \else
    \cutSpaceAfterCaption
    \fi
\end{figure}
\fi
\section{Related work}\label{sec:related}
\ifdefined\LARGEFIGURES
\begin{figure*}[t]
    \centering{
        \subfloat[Heavy Hitters Recall]{\includegraphics[width =\SIGMETRICSFigureSize]
            {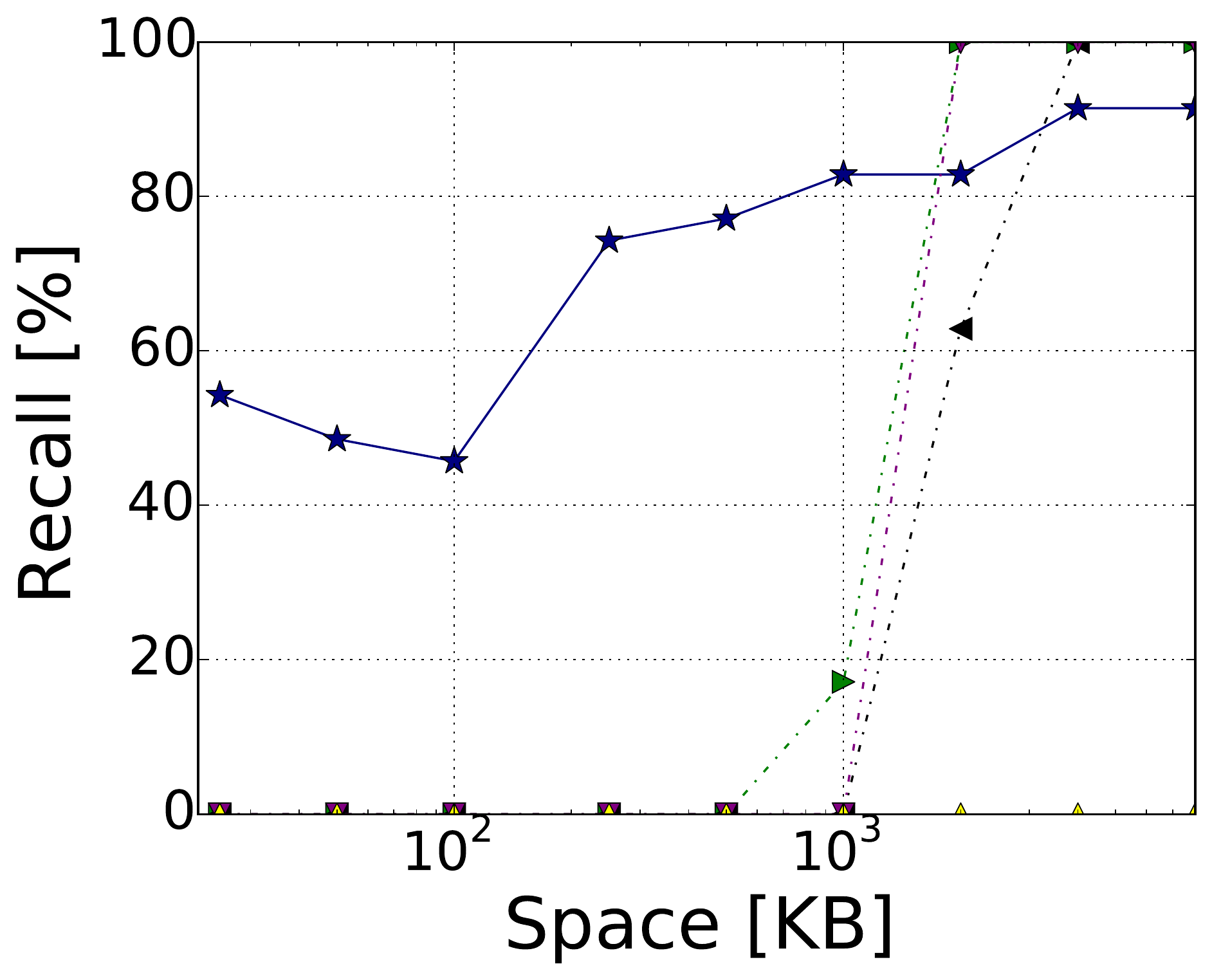}} 
        \subfloat[Heavy Hitters Precision]{\includegraphics[width =\SIGMETRICSFigureSize]
            {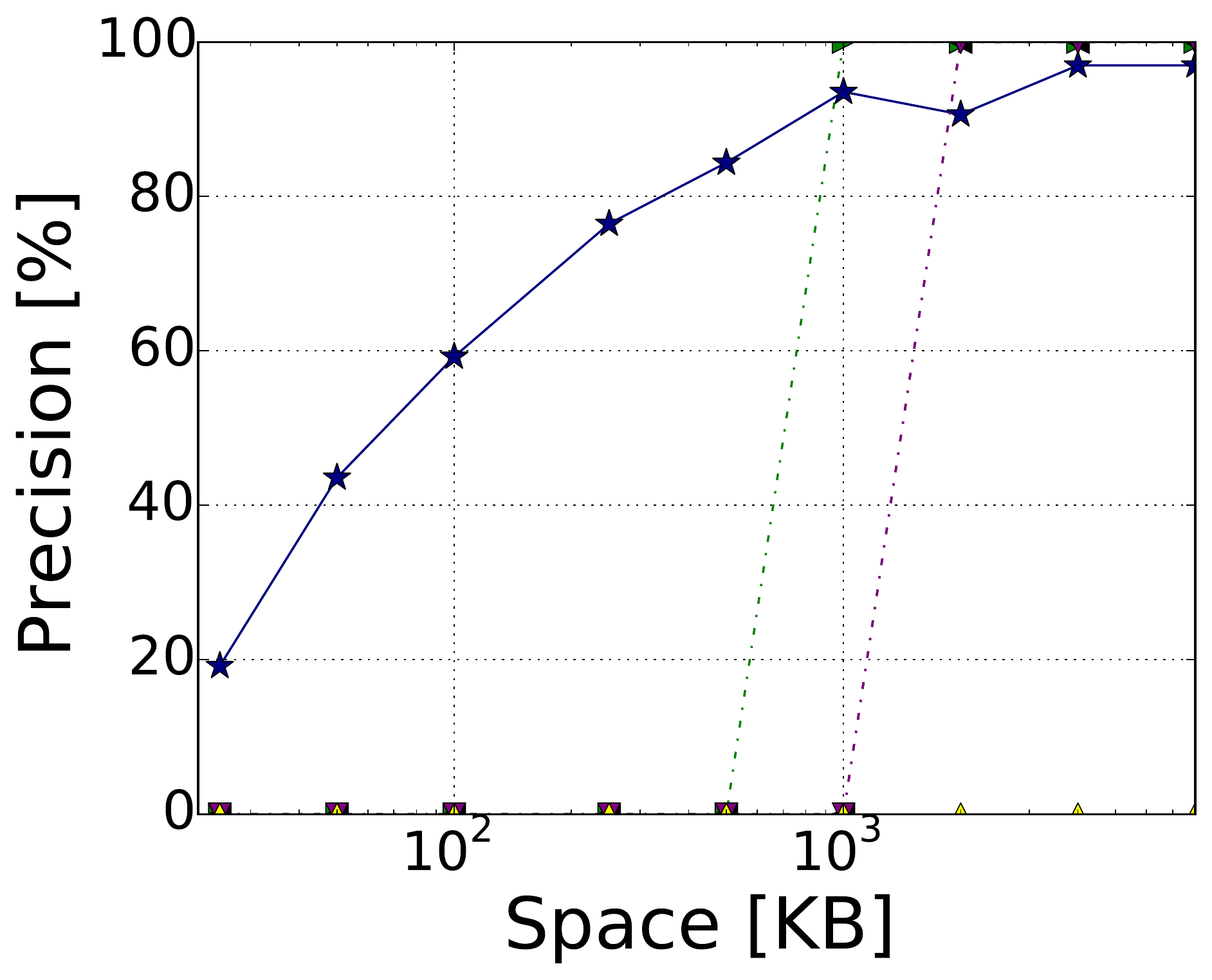}} \\
        \subfloat[Superspreaders Recall]{\includegraphics[width =\SIGMETRICSFigureSize]
            {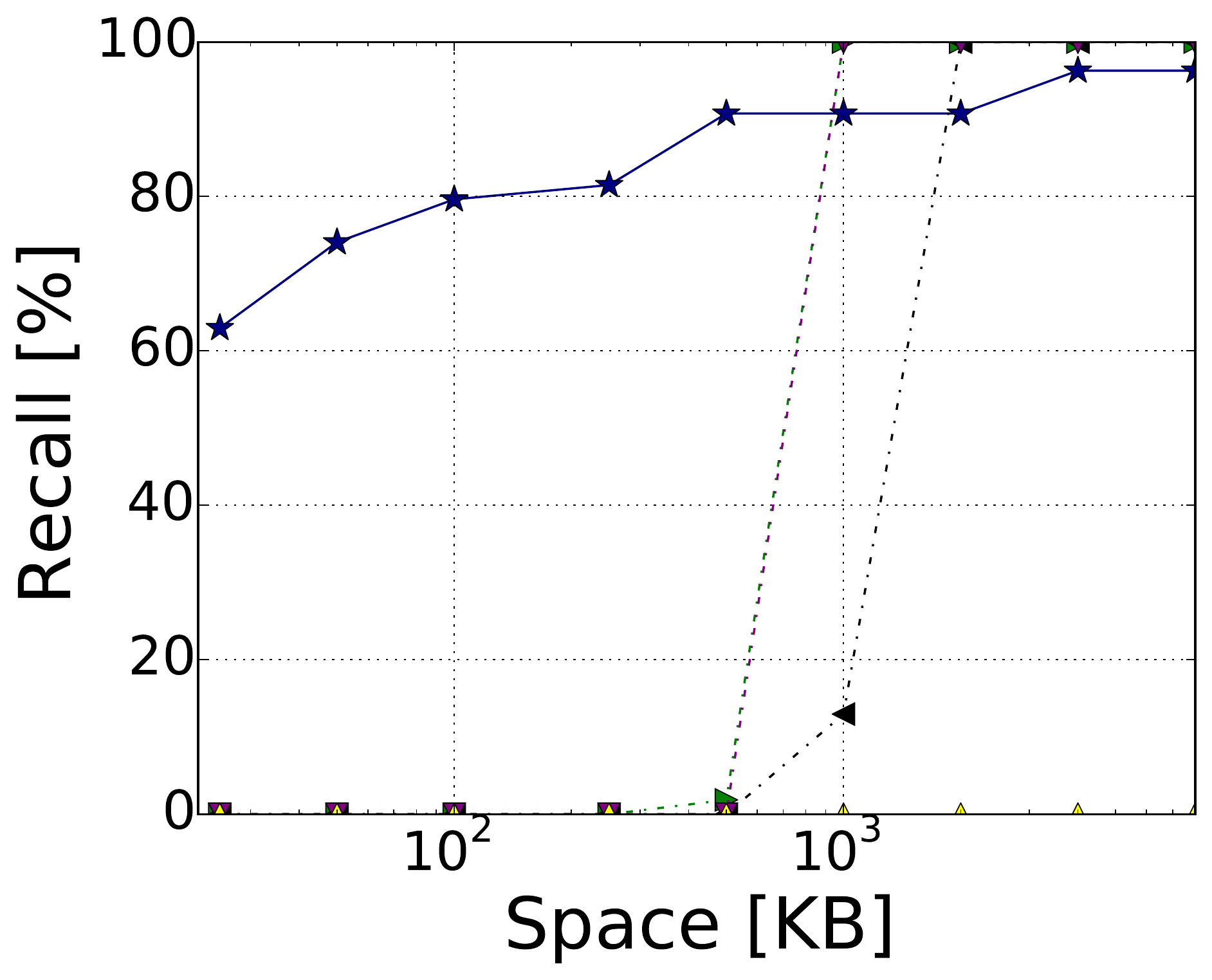}} 
        \subfloat[Superspreaders Precision]{\includegraphics[width =\SIGMETRICSFigureSize]
            {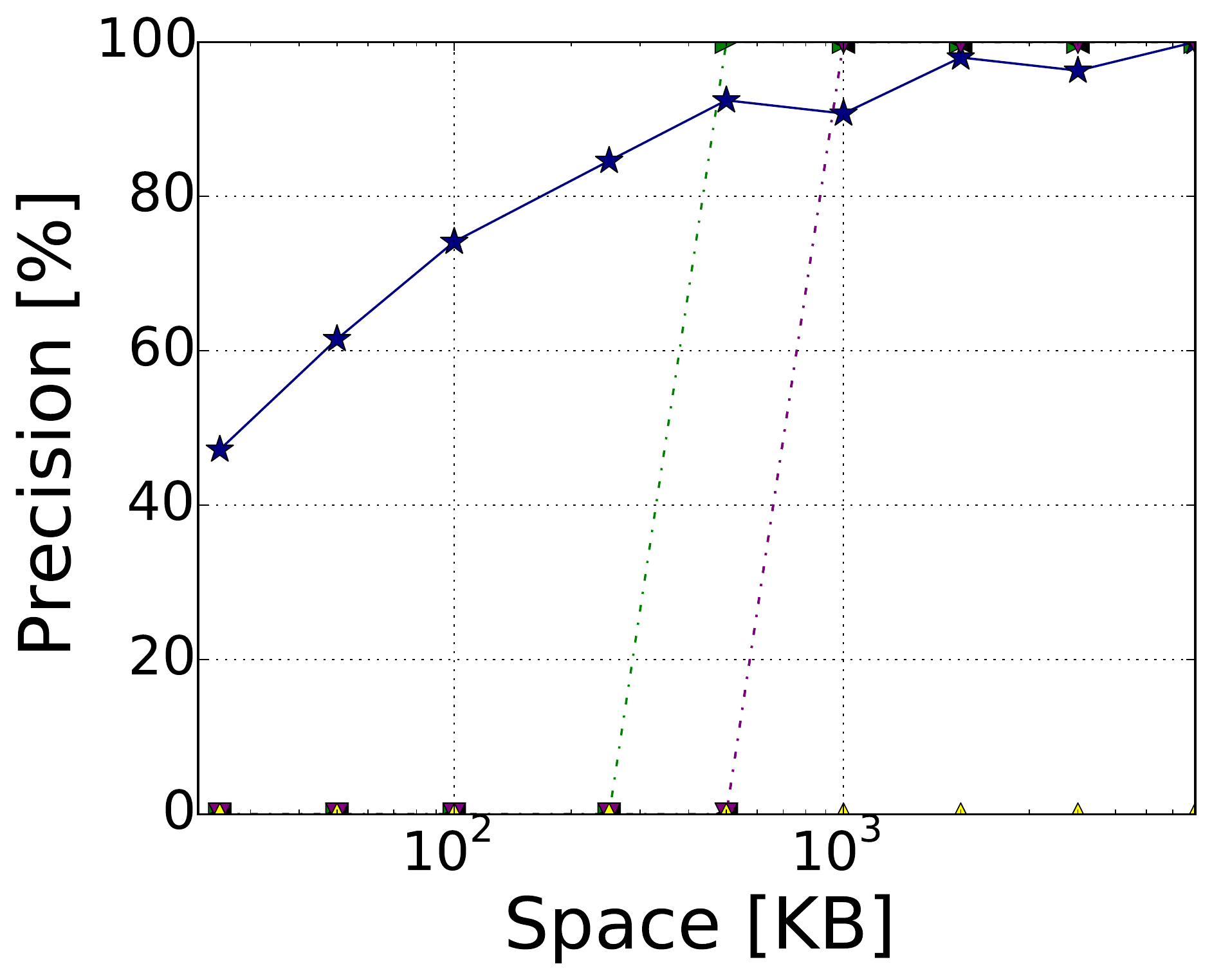}}             
    
    \includegraphics[width=1.3\columnwidth]{Figures/legend.png}}
    \caption{\label{fig:precisionRecall} The precision and recall breakdown of the algorithms for the HH and SS tasks. FlowRadar always has a perfect precision as any flow it decodes is fully~accurate.}
    \cutSpaceAfterCaption
\end{figure*}
\else
\begin{figure}[]
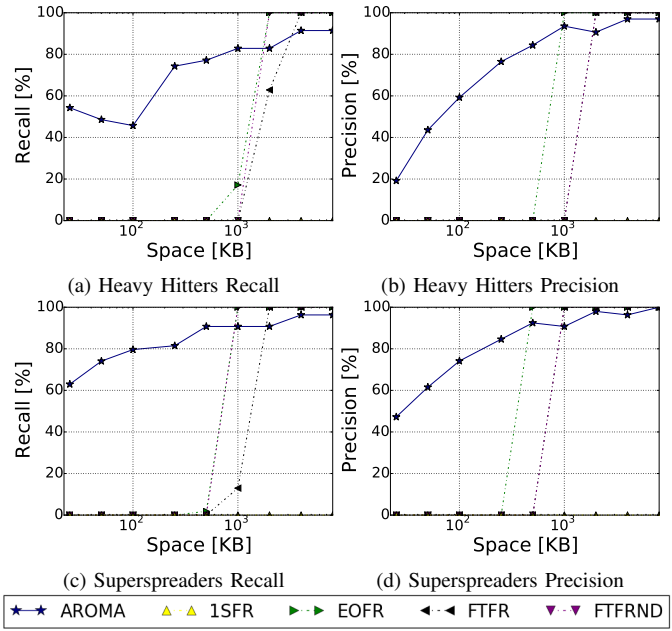

    \ifdefined\arXiv
    \else
    \vspace{-0.5cm}  
    \fi
    \centering{
        \subfloat[Heavy Hitters Recall]{\includegraphics[width =0.49\columnwidth]
            {Figures/recall_NY2018_HH}} 
        \subfloat[Heavy Hitters Precision]{\includegraphics[width =0.49\columnwidth]
            {Figures/precision_NY2018_HH}} \\\vspace{-4mm}
        \subfloat[Superspreaders Recall]{\includegraphics[width =0.49\columnwidth]
            {Figures/recall_NY2018_SS}} 
        \subfloat[Superspreaders Precision]{\includegraphics[width =0.49\columnwidth]
            {Figures/precision_NY2018_SS.pdf}}             
    \\
    \includegraphics[width=\columnwidth]{Figures/legend.png}}
    \caption{\label{fig:precisionRecall} The precision and recall of the algorithms for the HH and SS tasks. FlowRadar always has a perfect precision as any flow it decodes is fully~accurate.}
    \ifdefined\arXiv
    \else
    \cutSpaceAfterCaption
    \vspace{-0.3cm}  
    \fi
\end{figure}
\fi
We now survey network-wide techniques, as well as work that is related to the methods used in this work.

\textbf{Packet Marking: }
The work of~\cite{HHTagging} suggests marking measured packets by exploiting unused bits in the IP header. That way, they measure each packet once regardless of the number of measurement switches it traverses. 
However, this simple and effective method implicitly restricts measurement switch deployment. Intuitively, the unused bits need to be cleared before they enter our network. Otherwise, the method may fail due to a proprietary use of these \mbox{bits in other networks.} 

\textbf{Single per-flow Path Solutions: }
FlowRadar~\cite{FlowRadar}, EverFlow, and Trajectory sampling~\cite{Everflow,TrajectorySampling} assume that each flow is routed on a single path. 
The single path solutions cannot handle routing changes, multicast, and  Multipath routing~\cite{rfc6824} which are important in modern networks. 


\textbf{Flow Sampling Techniques: }
Several solutions have been proposed for flow sampling~\cite{cSamp,flowSampConstraints}. Specifically, in~\cite{cSamp}, the authors present cSamp, a flow sampling method that performs hash-based packet selection to coordinate between the measurement switches. cSamp performs network-wide monitoring by distributing responsibilities across the measurement switches in the network. The framework is responsive to routing, topology and network dynamics and shifts the responsibilities according to the network changes. In contrast, \sysName achieves network-wide uniform flow sampling without assigning specific responsibilities and therefore is not affected by the network~dynamics.

\textbf{Other routing oblivious solutions:  }\label{sec:others}
The BEFMR18 software-based algorithm~\cite{NetworkWideANCS} performs network-wide measurements through uniform packet sampling, using the same routing-oblivious assumption as this work. 
\ifdefined\arXiv
\vbox{\noindent However, BEFMR18 is not compatible with the architecture of PISA programmable switches, and it does not support flow~sampling.} 
\else
\noindent However, BEFMR18 is not compatible with the architecture of PISA programmable switches, and it does not support flow sampling.
\fi

\section{Conclusion}
\label{sec:discussion}
We introduced AROMA, a network-wide measurement infrastructure that enables network-wide flow and packet sampling in PISA switches. 
AROMA does not make any assumptions regarding routing and is flexible with respect to the placement of the measurement \mbox{switches in the network. }

\ifdefined\arXiv
Our work focuses on network-wide measurements on high-performance programmable network switches. 
We introduce a new measurement infrastructure that enables network-wide sampling in the data plane while considering the programming limitations of the switches. Moreover, this solution does not not make any assumptions regarding routing and is very flexible with respect to the placement of the NMPs. 
\else
\fi

We proved formal accuracy guarantees and demonstrated the ability to perform a variety of network measurement tasks. We evaluated AROMA through simulations with different topologies, per-switch memory, and measurement length. 
AROMA outperforms uniform sampling and that it allows accurate measurements in memory-constrained configurations where the previous works are inapplicable. 

AROMA's novelty extends beyond programmable switches. Specifically, it is the first technique to perform flow sampling without assumptions on the workload or coordination between the switches. 
Interestingly, it also has advantages in software implementation; specifically, it improves the update time of the existing (software) network-wide packet sampling technique~\cite{NetworkWideANCS} from logarithmic to a~constant. 

\textbf{Acknowledgements:}\quad{} This work is supported in part by the CNS1834263 grant, the Ben-Gurion University Cyber-Security Research Center, and the Zuckerman Institute.

\bibliographystyle{abbrv}
\bibliography{References}
\end{document}